\documentclass[singlecolumn, useAMS, usenatbib, graphicx]{mnras}
\usepackage[dvips]{graphicx}
\usepackage{amsmath,amsfonts}
\usepackage[retainorgcmds]{IEEEtrantools}
\usepackage{longtable}
\newcommand{\sub}[1]{\ensuremath{_{\mbox{\scriptsize#1}}}}

\title[Observable features of protoplanetary discs]{Observable scattered light features from inclined 
and non-inclined planets embedded in protoplanetary discs}

\author[D. L. Kloster et al.]{
Dylan L. Kloster,$^{1}$
M. Flock$^{2,3}$
\\
$^{1}$University of Wyoming, 1000 E. University, Dept 3905, Laramie, WY 82071, USA\\
$^{2}$Jet Propulsion Laboratory, California Institute of Technology, Pasadena, California 91109, USA\\
$^{3}$Max Planck Institute for Astronomy, K\"{o}nigstuhl 17, D-69117 Heidelberg, Germany}

\begin{document}

\date{}

\maketitle


\begin{abstract}
 Over the last few years instruments such as VLT/SPHERE and Subaru/HiCIAO have been
able to take detailed scattered light images of protoplanetary discs.  Many of the features observed 
in these discs are generally suspected to be caused by an embedded planet, and understanding the cause of these features requires detailed  theoretical models.  
In this work we investigate disc-planet interactions   
using the PLUTO code to run 2D and 3D hydrodynamic simulations of protoplanetary discs 
with embedded 30 M$_{\earth}$ and 300 M$_{\earth}$ planets on both an inclined ($i = 2.86^{\circ}$) and non-inclined orbit, using an $\alpha$-viscosity of $4 \times 10^{-3}$.  
We produce synthetic scattered-light images of these discs
at \emph{H-band} wavelengths using the radiative transfer code RADMC3D.
We find that while the surface density evolution in 2D and 3D simulations of inclined and non-inclined planets remain fairly similar, their observational appearance is remarkably different.
Most of the features seen in the synthetic \emph{H-band} images are
connected to density variations of the disc at around 3.3 scale heights above and below the midplane, which emphasizes the need for 3D simulations.  
Planets on sustained orbital inclinations disrupt the disc's upper-atmosphere and produce radically different observable 
features and intensity profiles, including shadowing effects and intensity variation in the order of 10-20 times the surrounding background. The vertical optical depth to the disc midplane for \emph{H-band} wavelengths 
is $\tau \approx 20$ in the disc gap created by the high-mass planet. We conclude that direct imaging of planets embedded in the disc remains difficult to observe, even for massive planets in the gap. 
\end{abstract}


\section{Introduction}\label{intro}

Planets are formed in the accretion discs of interstellar gas 
and dust around young stellar objects (YSOs).  
As these discs are believed to be regions of early planet formation and incubation,
they are commonly referred to as protoplanetary discs.
The planet formation process itself is not directly observable due to the distances
to these YSOs, the relatively small sizes of protoplanets, and that the 
newly forming planets are obscured within their host disc.  Therefore, one of the most observationally 
accessible ways of understanding planet formation is by studying 
protoplanetary discs via multi-wavelength
observations.  By comparing the observations with theoretical models,
we can better understand these discs themselves as planet-forming regions,  
as well as how the presence of a newly forming planet affects their dynamics
and evolution.

Accretion discs have been studied for decades, with theoretical models 
developed for their structure and evolution 
\citep[e.g.,][]{1973A&A....24..337S, 1981ARA&A..19..137P}.  
Later, properties of passive accretion discs specifically
around YSOs, such as scale height,
temperature profiles, and vertical structure were modeled from theory and observations 
\citep{1987ApJ...323..714K, 1997ApJ...490..368C, 1998ApJ...500..411D}.  
These model-disc profiles, combined with more readily available computational clusters,
allowed theorists to explore how  planets interact gravitationally with their host disc, 
and how the dynamics of the disc evolve with time via two-dimensional and three-dimensional
hydrodynamic (HD) simulations
\citep[e.g.,][]{2003MNRAS.341..213B, 2004ApJ...612.1152V, 2007MNRAS.377.1324C, 2008MNRAS.387..387E}.
These simulations revealed that planets can greatly affect their host disc, 
creating features such as spiral arms and density gaps due to displaced material 
in the orbital vicinity of the planet. 
See reviews by \cite{2011ARA&A..49..195A}, \cite{2011ARA&A..49...67W}, and \cite{2017arXiv170707148K} for a more complete 
description of protoplanetary discs.

Recent technological advances have shown evidence
that there may indeed be material gaps in protoplanetary discs seen in near-infrared (near-IR) wavelength observations in the form of dark rings, e.g. in TW Hya seen with Subaru \citep{2015ApJ...802L..17A} and 
in HD 169142 seen with VLT/SPHERE/ZIMPOL \citep{2018MNRAS.474.5105B}.
Other features like large-scale spiral-arm structures
have also been seen in near-IR observations of protoplanetary discs,  
such as SAO 206462 \citep{2013A&A...560A.105G} and MWC 758 \citep{2015A&A...578L...6B}. 
Spirals have been accompanied by shadowing effects in HD 100453 \citep{2017A&A...597A..42B},
and possible vortices and spirals caused by a planet in the HD135344B transition disc \citep{2016ApJ...832..178V}, as well as 
shadows cast on HD 135344B from multiwavelength VLT/SPHERE polarimetric differential imaging \citep{2016A&A...595A.113S}.

As the observations merely capture a snap-shot in time, and the planets themselves 
are not visible, a common strategy for protoplanetary disc simulations is to vary 
conditions such that the observable features can be recreated, then offer these conditions as 
a possible description of the protoplanetary system.  The dark  and bright rings commonly seen 
in observations have successfully been recreated in simulations from planets displacing material within their orbit and creating material 
gaps in the disc \citep[e.g.,][]{2015MNRAS.451.1147J, 2017ApJ...835..146D}.
\cite{2018A&A...617A..44K} and \cite{2018A&A...617L...2M} recently
observed the first stellar companion within the gap of a still-intact transitional 
disc, giving the results from these models derived via disc-planet simulations new weight.
However, recent HD simulations have been able to produce multiple gaps in a disc with only a 
single embedded planet \citep{2014ApJ...785..122Z, 2017ApJ...843..127D, 2017ApJ...850..201B}
suggesting it is possible that not every observed gap is due to an individual planet.

Synthetic images from hydrodynamic simulations have also been able to create observed spiral features in discs, usually as a result
of the presence of a massive planet or other stellar 
companion \citep{2015ApJ...809L...5D, 2015ApJ...812L..32D, 2016ApJ...816L..12D}.
These kinds of simulations have been able to successfully reproduce the spiral 
arms observed in systems such as SAO 206462 with a 10-15 M$_J$ companion \citep{2016ApJ...819..134B},
and HD 100453 using a 0.2 M$_{\odot}$ companion, which is actually observed 
in the system \citep{2018ApJ...854..130W}.  Although theories that do not require the presence 
of a planet have been suggested to explain spiral arms (e.g., disc shadowing \citep{2016ApJ...823L...8M}),   
the success of disc-companion simulations in recreating observable features 
of specific systems demonstrate that planets can visibly affect 
their host disc.  

These kinds of two- and three-dimensional simulations, combined with follow-up synthetic images,
are essential in understanding protoplanetary systems because of the relatively 
small number of clearly observable discs.  In most cases limitations in resolution  or obscuration
make directly observing the object responsible for producing the spiral-arm 
or gap features observed in some of these discs impossible or impractical.       
Simulating disc-planet interactions present the possibility of identifying 
the causes of these observed features, and models derived from simulations 
also have the potential to estimate planet properties based on its effects 
on the disc.   

The simulations used to model the observed systems discussed above used a simplifying assumption
that the orbit of the planet is aligned with the midplane 
of the disc, ignoring the possibility of planetary interactions 
causing orbital inclinations. Although it is unknown whether this kind of misalignment can be produced or sustained 
while the gas of the disc is still present, the idea of non-ideal planet-disc interactions (i.e.~planets with non-zero
orbital inclinations and eccentricities) has been of interest for years
\citep{2009ApJ...705.1575M, 2013A&A...555A.124B, 2017A&A...598A..70S, 2017MNRAS.468.4610C}, 
with more recent studies exploring changes in observable features of discs
containing inclined planets, such as \cite{2018MNRAS.475.3201A} and \cite{2019MNRAS.483.4221Z}.  

Here we continue this exploration by comparing
two- and three-dimensional simulations of protoplanetary 
discs with an embedded planet.  In the case of the three-dimensional simulations, 
we simulate this disc-planet evolution for planets with orbits aligned 
and mis-aligned with the disc midplane.  In all cases we produce 
synthetic images of how these simulated discs would be 
observed in near-infrared wavelengths using a Monte-Carlo radiative transfer method. 
We present our methods and initial conditions for our
hydrodynamic simulations in \S\ref{numMethods}.  Specific parameters used for the discs 
and the results of the HD simulations are given in \S\ref{HD} and 
the synthetic images as results from our radiative transfer simulations
in \S\ref{synthImages}.  In \S\ref{discussion}
we discuss these results, explore how orbital inclination damping would 
affect observable features in \S\ref{damping}, and derive conclusions 
from these results in \S\ref{summary}.

\section{Numerical Methods}\label{numMethods}

\subsection{Hydrodynamics}\label{HDs}

We model the dynamics and evolution of protoplanetary discs as 
a fully 3-dimensional, viscous, locally isothermal (no energy equation), hydrodynamic 
system governed by the equations of conservation
of mass and momentum 
\begin{eqnarray}
\frac{\partial \rho}{\partial t} + \nabla \cdot \rho \mathbf{v} & = & 0\label{consMass}\\*
\frac{\partial \mathbf{m}}{\partial t} + \nabla \cdot \left( \mathbf{mv} + \
P\mathbf{I}\right) & = & \
-\rho \nabla\Phi + \nabla\cdot\mathbf{\Pi} \label{consDen}
\end{eqnarray}
where $\rho$ is the mass density, $\mathbf{m} = \rho \mathbf{v}$ is the momentum density, 
$\mathbf{v}$ is the velocity, $P$ is the thermal pressure, $\mathbf{I}$ is the 
unit tensor. 
$\Phi$ is the 
softened gravitational
potential due to the central star and the planet, which is
\begin{equation}
\Phi = -\frac{GM_*}{r} - \frac{GM_p}{\sqrt{(\left|\mathbf{r} - \mathbf{r_p}\right|)^2
+\epsilon^2}},\label{planetPhi}
\end{equation}
where $r = \left|\mathbf{r}\right|$ is the
distance to the star and $\left|\mathbf{r} - \mathbf{r_p}\right|$ is the distance 
to the planet located at $\mathbf{r_p}$.
We use a spherical 
coordinate system, with radial, polar, and azimuthal variables as 
$\mathbf{r}$ = ($r,\theta,\phi$), 
unless otherwise specified since it is especially suited to describe the gravitational 
potentials of 
a planet with mass $M_p$ rotating around a central star of mass $M_*$.  The softening 
parameter is some fraction of
the Hill radius, $\epsilon = kr\sub{H}$, where
\begin{equation}
r\sub{H} \equiv r_p\left(\frac{M_p}{3M_*}\right)^{1/3}\label{rHill}
\end{equation}
and we use a value of $k=0.3$ throughout the series of simulations.
Viscous stresses are handled via the viscous stress tensor, $\mathbf{\Pi}$, with components
given by
\begin{equation}
(\Pi)_{ij} = \mu_1\left(\frac{\partial v_i}{\partial x_j} + \
\frac{\partial v_j}{\partial x_i}\right) + \
\left(\mu_2 - \frac{2}{3}\mu_1\right) \nabla \cdot \mathbf{v} \delta_{ij}\label{viscTensor}
\end{equation}
where $\mu_1$ and $\mu_2$ are the coefficients of shear and bulk viscosity.  We use the typical $\alpha$-viscosity
prescription of \cite{1973A&A....24..337S} defined as 
\begin{equation}
\nu = \alpha c_s H\label{alpha}
\end{equation}
where $\nu$ is the kinematic turbulent viscosity, which is related to the 
coefficient of shear viscosity in Eqn.~(\ref{viscTensor}) as 
\begin{equation}
\mu_1 = \nu\rho\label{viscCoeff}
\end{equation}
and assume a coefficient of bulk viscosity as $\mu_2 = 0$, typical
for low-density density gases not experiencing hypersonic compression or expansion.

In order to solve the conservation equations (\ref{consMass}) and (\ref{consDen}), we use 
the numerical code PLUTO v4.2 \citep{2007ApJS..170..228M}, a modular, finite volume, 
shock-capturing code.   We choose a piecewise total variation diminishing 
linear reconstruction
integrator, which is 2nd order accurate in space, and 
specified the use of a monotonized central difference limiter (\verb+MC_LIM+).
We applied a 2nd order Runge-Kutta 
time integrator, a Courant-Friedrichs-Lewy (CFL) number
of 0.3, and in all cases we ran PLUTO with the FARGO 
module to save computational time.  
 
All simulations use an isothermal equation of state
standard in the 
PLUTO code, which defines the relation between gas density and pressure as 
\begin{equation}
P = c_s^2 \rho \label{eos}
\end{equation}
where $c_s$ is the sound speed of the gas. 
Assuming an ideal gas, the relation between $c_s$ and
gas temperature, $T_g$, is
\begin{equation}
c_s^2 = \frac{k_b T_g}{\mu m_p}\label{sSpeed}
\end{equation}
where $k_b$ is the Boltzmann constant, $\mu$ is the mean molecular weight,
and $m_p$ is the mass of a proton.   For a protoplanetary disc, 
a typical used value for the mean molecular weight is $\mu = 2.3$, to represent
a gas composition consists primarily of molecular hydrogen, plus a smaller 
amount of heavier elements.

For a thin disc in hydrostatic and thermal equilibrium that is heated only
by radiation from a central host star, the gas temperature profile can be approximated    
as being vertically isothermal for the disc interior, 
depending only on the distance from the host
star \citep{1987ApJ...323..714K}.  
This approximation is consistently used in similar hydrodynamic 
and magnetohydrodynamic simulations of protoplanetary discs 
\citep{2011ApJ...736...85U, 2013MNRAS.435.2610N}, 
and in this case the gas temperature is proportional
to a power-law as $T_g \propto (r\sin\theta)^{q}$, which from Eqn.~(\ref{sSpeed})
implies $c_s^2 \propto (r\sin\theta)^{q}$. 
We modified the PLUTO code so that this sound-speed profile could be included 
in the equation of state.

\subsection{Initial and Boundary Conditions}\label{initConds}

To derive the initial conditions it is useful to use a cylindrical coordinate system, 
with radial, azimuthal, and vertical variables as $\mathbf{R}$ = ($R,\phi,z$), 
where spherical and cylindrical radial coordinate variables are related as $R = r\sin\theta$.
These conditions are derived from Eqn.~(\ref{consDen})
for a non-viscous
circumstellar disc ($\mathbf{\Pi} = 0$)  without a planet,
\begin{equation}
\frac{\partial \mathbf{v}}{\partial t} + \left(\mathbf{v} \cdot \nabla \right) \mathbf{v}\
=  -\frac{1}{\rho}\nabla P - \nabla \Phi_*,\label{simpleCM}
\end{equation}
where $\Phi_* = GM_*/(R^2+z^2)^{1/2}$ is the gravitational potential 
due only to the central star. 
In both models the disc is in hydrostatic equilibrium in the radial 
and vertical directions, has a mid-plane density of
\begin{equation}
\rho\sub{c}(R) = \rho_0\left(\frac{R}{R_0}\right)^p\label{cDen}
\end{equation}
and a radially dependent gas sound-speed of
\begin{equation}
c_s^2(R) = c_0^2\left(\frac{R}{R_0}\right)^q\label{cSpeed}
\end{equation}
where $R_0$ is the initial orbital radius of the planet, $\rho_0$ is
the mid-plane density at $R_0$, $c_0^2$ is the square of the sound speed at $R_0$, 
and $p$ and $q$ are power law indices. 
Separating Eqn.~(\ref{simpleCM}) into its vertical and
radial components
\begin{eqnarray}
0 & = & \frac{1}{\rho}\frac{\partial P}{\partial z} + \
\frac{GM_*}{(R^2+z^2)^{3/2}}z\label{vert}\\*
R\Omega^2 & = & \frac{1}{\rho}\frac{\partial P}{\partial R} + \
\frac{GM_*}{(R^2+z^2)^{3/2}}R.\label{rad}
\end{eqnarray}

Initial profiles for $\rho(R,z)$ and $\Omega(R,z)$ come from solving 
Eqs.~(\ref{vert}) and (\ref{rad}) 
directly, 
with solutions from \cite{2013MNRAS.435.2610N}:
\begin{eqnarray}
\rho(R,z) & = & \rho_0\left(\frac{R}{R_0}\right)^p \
\exp\left(\frac{GM}{c_s^2}\left[\frac{1}{\sqrt{R^2 + z^2}} - \
\frac{1}{R}\right]\right)\label{initRho}\\*
\Omega(R,z) & = & \Omega_k\left[(p+q)\left(\frac{H}{R}\right)^2 + (1+q) - \
\frac{qR}{\sqrt{R^2 + z^2}}\right]^{1/2}\label{Omega}
\end{eqnarray}
where $\Omega_k = \sqrt{GM/R^3}$ is the Keplerian angular velocity at radius $R$,
and $H \equiv c_s/\Omega_k$ is 
the thermal scale height of the disc.
The evolution of our planet's orbit is determined by parameterized equations 
in Cartesian coordinates 
\begin{eqnarray}
x_p(t) & = & -R_0\cos(\Omega_k t)\\*
y_p(t) & = & -R_0\sin(\Omega_k t)\cos(i)\\*
z_p(t) & = & -R_0\sin(\Omega_k t)\sin(i)\label{z_t}
\end{eqnarray}
where $t$ is time, and $i$ is the angle of inclination 
of the planet with respect to the midplane of the disc.
Setting $R_0 = 1$, this 
corresponds to the initial location of the planet being at 
$\mathbf{r_0} = (r_0, \theta_0, \phi_0) = (1, \pi/2, \pi)$
in our simulation units.  The location of the star was 
held fixed at the origin.  

We use a computational domain of
$r \in [0.3, 3.5]$, 
$\theta \in [1.3258, 1.8158]$, and $\phi \in [0,2\pi]$, 
with a uniform grid of 
$(N_r,N_{\theta},N_{\phi}) = (400,98,786)$ cells, which
allows us 
to resolve one vertical scale height with 10 cells 
at the orbital radius of the planet.
The parameters used for the basic setup of all our hydrodynamic 
simulations are summarized in Table \ref{parameters}.
  
We allow the planet to grow to full mass 
over the course of 10 orbits so the disc in hydrostatic equilibrium is not suddenly 
disrupted by the appearance of a gravitational potential.  Azimuthal boundary conditions 
are periodic, and we use a zero-gradient 
outflow condition at the polar boundaries which forces ghost-cell velocity values 
to zero if they are directed back into the disc.   To prevent mass loss, ghost cells at the radial boundaries 
are fixed at initial conditions. 

\begin{figure}
\includegraphics[width=\columnwidth]{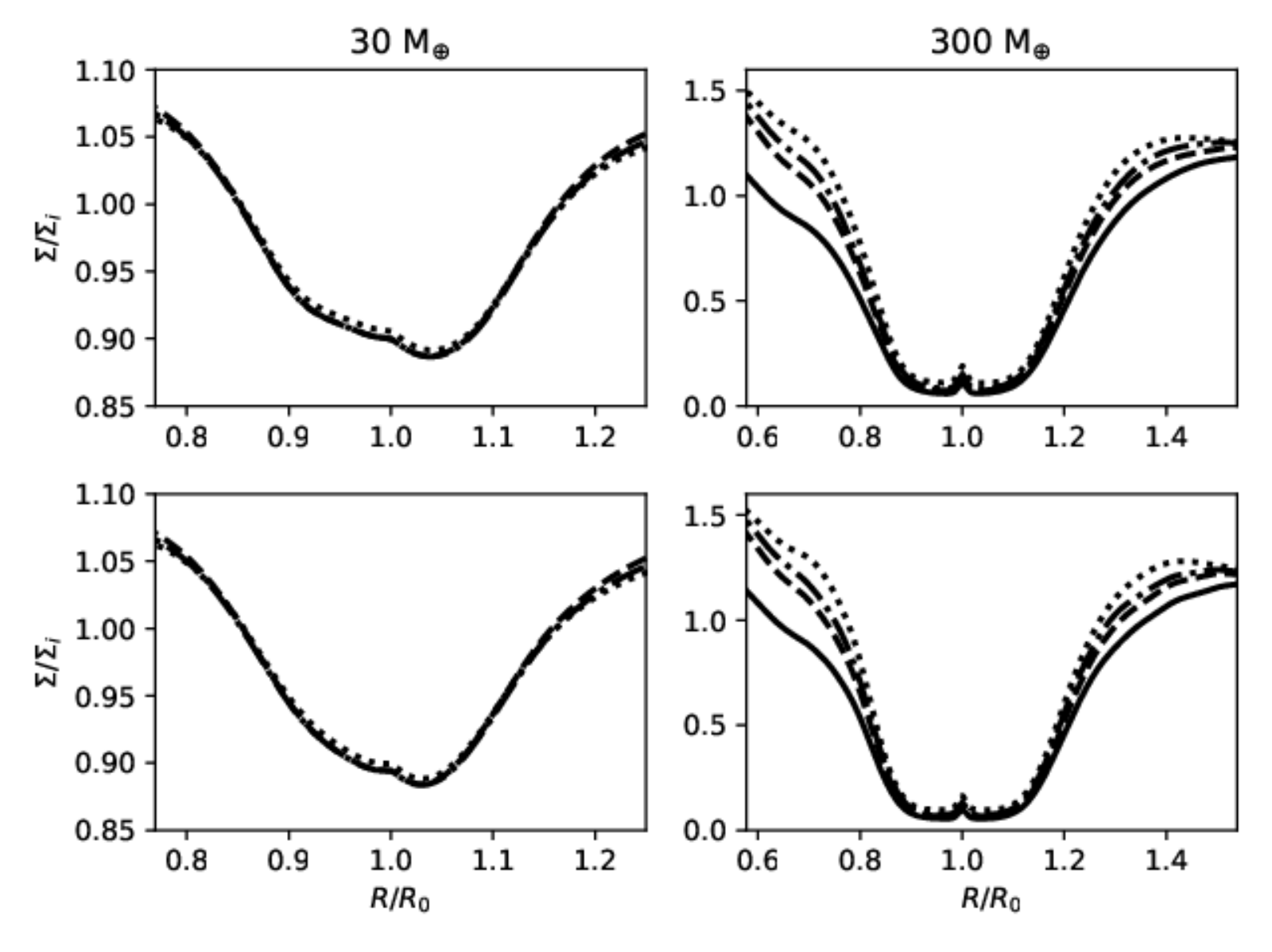}
\caption{Time evolution of normalized disc surface density for simulations
involving planets with orbits aligned with the disc
midplane (Top) and misaligned with the disc (Bottom).  Dotted ($\cdot\cdot$) lines 
show the state of the disc at $t = 200$ orbits, Dash-dotted ($-\cdot-$) lines 
are at $t=300$ orbits, dashed ($--$) lines are at $t = 400$ 
orbits, and the solid lines ($-$) are at $t = 1000$ orbits.  The depth 
of the dips in surface density have essentially reached a steady state.}\label{fourPlots}
\end{figure}

\begin{figure*}
\includegraphics[width=\textwidth]{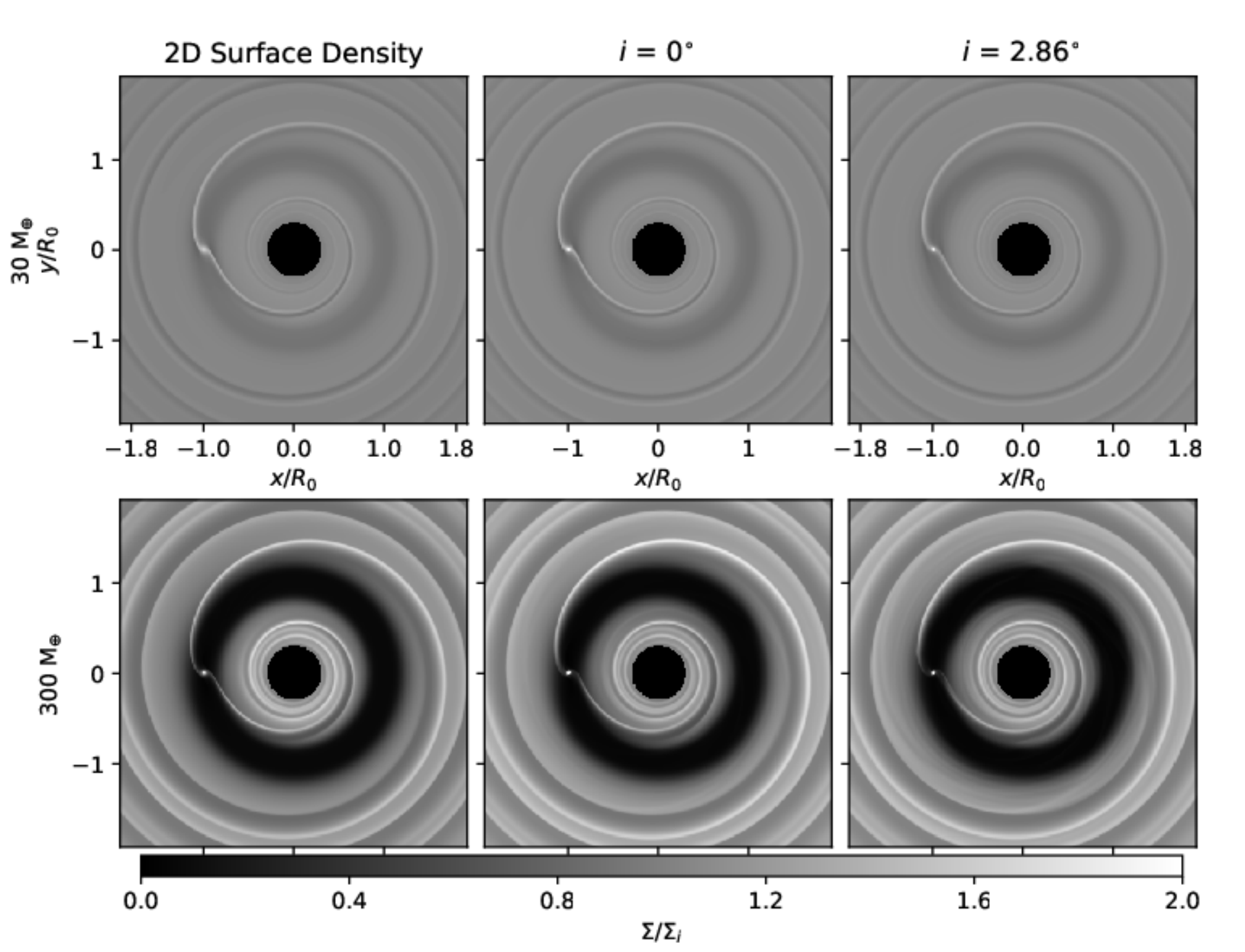}
\caption{Surface density maps of the 2D (Left), 3D non-inclined (Centre), and 
3D inclined (Right) planetary-orbit simulations with 30 M$_{\earth}$ (Top) and
300 M$_{\earth}$ (Bottom) planets.}\label{sdMaps}
\end{figure*}

\section{HD Simulation Setup \& Results}\label{HD}

\subsection{Simulation Setup}\label{simSetup}

We simulated a flared protoplanetary disc with a sound speed profile of 
$q = -1/2$ in Eqn.~(\ref{cSpeed}) to mimic a passive irradiated disc, 
which corresponds to a thermal scale-height profile of
\begin{equation}
 H = H_0 \left(\frac{R}{R_0}\right)^{5/4},
\end{equation}
where $H_0$ is the thermal scale-height at $R_0$.  
With this profile,
we used a central density index of $p = -7/4$ in Eqn.~(\ref{cDen})
in order to keep an initial surface density profile of 
\begin{equation}
\Sigma_i = \Sigma_0 \left(\frac{R}{R_0}\right)^{-1/2},
\end{equation}
where $\Sigma_0$ is the surface density of the unperturbed 
disc at $R_0$. We set $H_0 = 0.05 R_0$, which bounds the simulation
to 5$H$ above and below the midplane at the 
location of the planet.
In all simulations we use a constant $\alpha$-viscosity parameter
of $\alpha = 4.0\times 10^{-3}$.  

We studied the evolution of a 3D disc with an embedded 
planet whose orbit was aligned with the disc's midplane ($i=0^{\circ}$).
We considered planets with a mass of 30 M$_{\earth}$ (model \verb+3D30A+) and one of 
300 M$_{\earth}$ (\verb+3D300A+).   
Other previous works have shown the presence
of a planet will carve gaps in azimuthally averaged surface density
plots of protoplanetary discs \citep[e.g.,][]{2003MNRAS.341..213B, 2006Icar..181..587C}.
Exploratory test runs of our HD simulations revealed similar gaps, and we used
the depth of these gaps to determine the duration of the simulations.
We ran the 30 M$_{\earth}$ and 300 M$_{\earth}$ planet 
simulations until the azimuthally averaged surface densities in the vicinity 
of the planet (i.e. the depth of the gap) had essentially reached a steady-state. 
This corresponded to the 30 M$_{\earth}$ planet simulations being run for $\sim$200 orbits 
and the 300 M$_{\earth}$ planet simulations for $\sim$400 orbits, 
but we extended the 30 M$_{\earth}$ planet simulations out to
400 orbits and the 300 M$_{\earth}$ planet simulation out to 1000 orbits.
This is demonstrated in Figure \ref{fourPlots}, which shows the time evolution
of the azimuthally averaged surface density near the location of the planet 
at $t = 200, 300,$ and $400$ orbits of the 30 M$_{\earth}$ and 300 
M$_{\earth}$ planets, as well as at $t = 1000$ orbits of the 300 M$_{\earth}$
planet. This plot shows that the gap profile has converged.

For comparison, we ran simulations in which the disc evolved with a 
planet orbiting at a sustained slight incline with respect to the disc. 
We chose an inclination of $i= 2.86^{\circ}$, which corresponds to 
approximately one thermal scale-height above the midplane at the apex of 
the orbit.  We performed these inclined orbit simulations again using 
a 30 M$_{\earth}$ planet (\verb+3D30I+), as well as a 300 M$_{\earth}$ planet 
(\verb+3D300I+).  We also created 2-dimensional 
equivalent simulations of discs 
with embedded 30 M$_{\earth}$ (\verb+2D30+) and 300 M$_{\earth}$ (\verb+2D300+) planets for
400 and 1000 orbits, respectively, using the same
resolution and conditions described in \S\ref{initConds}, 
with an initial surface density
profile of $\Sigma_i = \Sigma_0\left(R/R_0\right)^{-1/2}$.
In these 2D simulations we used values of $p=-1/2$
and $z = 0$
for our initial conditions in Eqns.~(\ref{initRho}) and (\ref{Omega}), 
as well as an equation of state of $P = \Sigma c_s^2$.  
However, by their nature 2D simulations can not include information about the
vertical dynamics or the dynamics of the material above over below the 
planet.  Analysis and simulations of discs by \cite{2012A&A...541A.123M}
and \cite{2012A&A...546A..99K} showed that in order for the results from
2D simulations to be in agreement to their 3D counterparts, the softening 
parameter for the gravitational potential (Eqn. \ref{planetPhi}) must be 
increased to $\epsilon = 0.7H$ when self-gravity is not taken into 
consideration.  We therefore include this correction to our 2D simulations.

Model parameters for each of our simulations are summarized in Table \ref{models}.   

\begin{table}
\begin{center}
\caption{Physical parameters used for the general setup of hydrodynamic simulations.}
\label{parameters}
\begin{tabular}{lcc}
\hline
Parameter & Symbol & Value\\
\hline
Stellar mass & $M_*$ & M$_{\sun}$\\
Orbital radius & $R_0$ & 1.\\
Thermal scale-height at $R_0$ & $H_0$ & 0.05$R_0$\\ 
Alpha viscosity & $\alpha$ & $4\times 10^{-3}$\\
Number of grid cells & $N_r \times N_{\theta} \times N_{\phi}$ & $400 \times 98 \times 786$\\
Radial resolution & $\Delta r$ & $4H_0/25$\\
Polar resolution at $r_0$ & $r_0\Delta \theta$ & $H_0/10$\\
\hline
\end{tabular}
\end{center}
\end{table}

\subsection{300 M$_{\earth}$ Planet Results}\label{jupHDResults}

\begin{figure}
\includegraphics[width=\columnwidth]{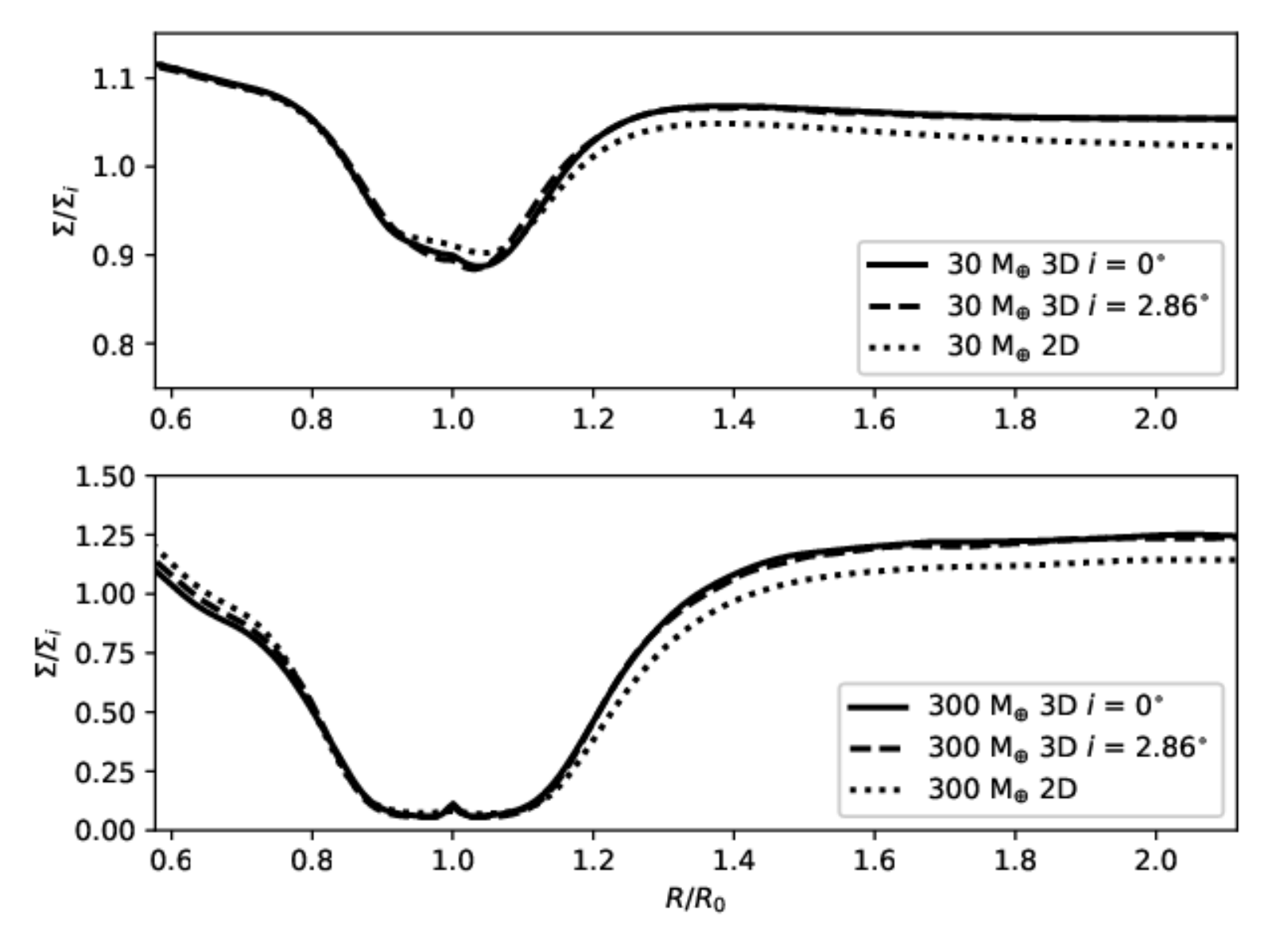}
\caption{Azimuthally averaged surface density profile, $\Sigma$, normalized by initial
surface density, $\Sigma_i$, for the 30 M$_{\earth}$ (Top) and
300 M$_{\earth}$ (Bottom) simulations for planet orbits aligned with the disc (solid, $-$),
misaligned with the disc (dashed, $--$), and 2D discs (dotted, $\cdot \cdot$).
Results from the 30 M$_{\earth}$ planet simulations are after 400 orbits, 
and after 1000 orbits for the 300 M$_{\earth}$ planet simulations.}\label{sdAvg}
\end{figure}

Final global surface density features for each simulation are presented 
as maps in Figure \ref{sdMaps}, as well as azimuthally averaged 
profiles in Figure \ref{sdAvg}.  In the case of the 300 M$_{\earth}$ planet 
simulations, the features produced from the disc-planet interaction were independent 
of the dimensionality of the simulation or whether the orbit of the planet was 
aligned with the disc. In all three cases the planet creates an annular low-density region in the 
orbital vicinity of the planet, as well as primary spiral-density arms which 
emanate inwards and outwards from the planet.  A secondary arm can also be seen
in each simulation inwards of the planet's orbit around $(x/R_0,y/R_0) = (0.75,0)$, 
along with the possibility of a tertiary near $(-0.5,0)$.  These various 
arms are expected as even tertiary arms have been seen in similar 
hydrodynamic simulations of discs with embedded high-mass planets \citep[e.g.,][]{2017ApJ...835...38D}. 

The similarities in simulations involving the 300 M$_{\earth}$ planet
are even more obvious when presented in Figure \ref{sdAvg} (Bottom).
As seen in similar HD simulations 
\citep[e.g.,][]{2003MNRAS.341..213B,2006MNRAS.370..529D,2011ApJ...736...85U}, 
the high-mass planet created a substantial gap in its vicinity.
The depth, width, and general shape of this gap is remarkably unaffected 
by the variations of each simulation.  
\cite{2015MNRAS.448..994K, 2015ApJ...806L..15K} contributed to developing a 
relationship between the depth of such gaps as
\begin{equation}
\frac{\Sigma\sub{min}}{\Sigma_i} = \frac{1}{1+ 0.04K},\label{preDepth}
\end{equation}
where $\Sigma\sub{min}$ is the surface density at the bottom of the gap,
\begin{equation}
K = \left(\frac{M_p}{M_*}\right)^2\left(\frac{H}{R}\right)^{-5}\alpha^{-1},
\end{equation}  
and, again, $\Sigma_i$ is the surface density of the unperturbed disc.

With this as a basis, \cite{2019MNRAS.483.4221Z} derived a relation for 
the expected gap depth caused by an inclined planet, using a new value 
for $K$ as 
\begin{equation}
    K = \left(\frac{M_p}{M_*}\right)^2\left(\frac{H}{R}\right)^{-5}\alpha^{-1}\frac{4}{\pi^2 (n+1) }\
\left[\mathcal{K}\left(\frac{n}{n+1}\right)\right]^2,
\end{equation}
where $n = (R_0{\sin}(i)/H)^2$,  
$\mathcal{K}(m)$ is the complete elliptic integral of the first kind, with $m$=$k^2$, 
and $k$ is the elliptic modulus. 

Values derived from Eqn. (\ref{preDepth}) predict 
that slight orbital inclinations of high-mass planets do not significantly affect gap 
depth, a feature that is also demonstrated in the simulations explored in this work.  
Although Figure \ref{sdAvg}
suggests that in all three cases depths of the gaps in surface density produced by a 300 M$_{\earth}$ 
planet are essentially identical, the numerical values of $\Sigma\sub{min}/\Sigma_i$  show variations 
between each other and the predicted values (Table \ref{models}).  
In each of our simulations, the depth of the 
gap is not quite as great as the predicted values, especially in the case of the 2D simulation.
When compared to results from other simulations \citep[e.g.,][]{2004ApJ...612.1152V, 2013ApJ...769...41D}, 
and \cite{2014ApJ...782...88F}, Eqn. (\ref{preDepth}) does fit rather well, even though there is 
a fairly large spread in this fit (See Figure 1 of \citep{2015ApJ...806L..15K}). 

As the results from our simulations involving a high-mass planet are within this spread, we can consider 
them to be in reasonable agreement with the predicted values, as well as other work.  However,
It should be noted that many of the simulations from \cite{2004ApJ...612.1152V}, \cite{2013ApJ...769...41D}, and 
\cite{2014ApJ...782...88F} were run for several thousand to 20,000 orbits to generate 
the values of their gap depths, and the limitations on the duration of these simulations 
due to their three-dimensional nature could be the cause of this discrepancy between 
simulation and prediction. While it is not likely that these small differences in gap depth
would drastically effect the observable features, it would be worth a single 
extended simulation dedicated to test this assumption.

\begin{table*}
\begin{center}
\caption{Summary of hydrodynamic disc models and parameters used in simulations.
Columns: 1) Model name, 2) Dimensions, 3) Planet mass, 4) Number of orbits at $R=R_0$,
5) Power-law index for initial $\rho$ (Eqn. \ref{initRho}),
6) Orbital inclination of planet (in degrees),
7) Gravitational potential softening parameter (Eqn. \ref{planetPhi}), 
8) Maximum gap depth from simulation,
9) Maximum predicted gap depth (Eqn. \ref{preDepth}).  
All simulations used a power-law index of $q=-1/2$  for $c_s^2$ (Eqn. \ref{cSpeed}).
}\label{models}
\begin{tabular}{ccccccccc}
\hline
Model & Dimensions & Planet Mass & Orbits & $p$ & $i$ & $\epsilon$ & Gap Depth & Predicted Depth\\
&&&&&&& $\Sigma\sub{min}/\Sigma_i$ & $\Sigma\sub{min}/\Sigma_i$\\
(1) & (2) & (3) & (4) & (5) & (6) & (7) & (8) & (9)\\
\hline
\verb+2D30+ & 2D & 30 M$_{\earth}$ & 400 & -1/2 & - & 0.7$H$ &  0.90 & 0.80\\
\verb+3D30A+ & 3D & $''$ & 300 & -7/4  & $0^{\circ}$ & 0.2$H_0$ & 0.89 & $''$\\
\verb+3D30A+ & 3D & $''$ & 400 & -7/4  & $0^{\circ}$ & $''$ & 0.89 & $''$\\
\verb+3D30I+ & 3D & $''$ & 300 & -7/4  & 2.86$^{\circ}$& $''$ & 0.88 & 0.85\\
\verb+3D30I+ & 3D & $''$ & 400 & -7/4  & 2.86$^{\circ}$& $''$ & 0.88 & $''$\\
\verb+2D300+ & 2D & 300 M$_{\earth}$ & 1000  & -1/2 & - & 0.7$H$ &  0.070 & 0.037\\
\verb+3D300A+ & 3D & $''$ & 300 & -7/4  & $0^{\circ}$ & 0.4$H_0$ & 0.083 & $''$\\
\verb+3D300A+ & 3D & $''$ & 700 & -7/4  & $0^{\circ}$ & $''$ & 0.061 & $''$\\
\verb+3D300A+ & 3D & $''$ & 1000 & -7/4  & $0^{\circ}$ & $''$ & 0.059 & $''$\\
\verb+3D300I+ & 3D & $''$ & 700 & -7/4  & 2.86$^{\circ}$ & $''$ & 0.057 & 0.053\\
\verb+3D300I+ & 3D & $''$ & 1000 & -7/4  & 2.86$^{\circ}$ & $''$ & 0.055 & $''$\\
\hline
\end{tabular}
\end{center}
\end{table*} 
 
\subsection{30 M$_{\earth}$ Planet Results}\label{e30HD}

Figures \ref{sdMaps} and \ref{sdAvg} also show results of our
hydrodynamic simulations involving the 30 M$_{\earth}$ planet after 400 orbits.
As in the 300 M$_{\earth}$ planet situation, the surface density maps show no obvious differences 
regardless of the dimensionality of the 
simulation or the alignment of the planet.
In each case 
spiral density waves are emitted radially inwards and outwards from the 
planet, causing a primary spiral arm coiling towards the inner 
boundary.  A secondary arm can also be seen near $(x/R_0,y/R_0) = (0,-0.75)$ in
Figure \ref{sdMaps} in all three cases, but these maps lack the tertiary arm seen in
the simulations involving a high-mass planet.  While the average surface 
density values are slightly lower in the 2D simulations radially outwards from the 
planet, the general shape of the profile is essentially the game.  One notable difference 
is the slightly deeper depths in averaged surface densities in the 3D cases 
compared to the 2D case (see Figure \ref{sdAvg} and Table \ref{models}), however 
this difference is quite small.

Values presented in Table \ref{models} do show a discrepancy between the 
gap depths from the simulations compared to those predicted from Eqn. (\ref{preDepth}), 
even though this ratio of $\Sigma\sub{min}/\Sigma_i$ remained constant from
$t = 200-400$ orbits (Figure \ref{fourPlots}).  Predicted values of gap depths 
are almost 10\% greater than what the 2D and 3D simulations show.  This is not a slight
discrepancy which could simply be attributed to scatter, as in the case of the 
high-mass planet simulation.

With our fixed boundary conditions, material moving radially outwards is not allowed 
to leave the system.  To test if this hindered the low-mass planet's ability to displace
material, causing the discrepancy between simulation and predicted gap depth, we  
tested our \verb+3D30A+ model using 
outflow conditions at the radial boundaries.  
With this set up, our 
gap depth was $\Sigma\sub{min}/\Sigma_i = 0.89$ at $t=200$ orbits, but dropped 
to $\Sigma\sub{min}/\Sigma_i = 0.88$ from $t=200-400$, slightly deeper than 
with the fixed boundary conditions.  We then extended the run out 
ever further to find that the gap depth increased very slowly where 
it finally reached a value of $\Sigma\sub{min}/\Sigma_i = 0.87$  at $t=800$.

While the outflow conditions do show the gap depth to continue to increase 
over time, we see diminishing returns similar to those seen in the 300 M$_{\earth}$
simulations; where several hundred orbits are required to increase the gap depth 
by a very small amount.  However, in the case of the high-mass planet simulation, 
the gap was essentially made by $\sim$400 orbits, and begin to slowly approach 
its predicted value of the course of the next several hundred orbits.  In the case of 
this low-mass planet, the gap is slowly increasing, but it is still no where near 
the predicted value of $\Sigma\sub{min}/\Sigma_i = 0.80$ after 800 orbits.  

Since the depth of this gap is slowing growing over time with the outflow 
bounds, it is possible it would match its predicted value after 
several thousand, or tens of thousands, of orbits, as mentioned in \S\ref{jupHDResults}.
However, is also possible that the gap depth is increasing because using 
outflow boundary conditions for a viscous disc result in an overall mass loss, 
with the total mass of the disc decreasing by 23\% over the first 400 orbits, 
and by 32\% after 800 orbits. This drastic loss of material in the inner 
and outer disc could affect 
observational features produced from these models, making much longer runs 
not viable.  Likewise, fixing the radial boundaries at initial conditions does 
slightly increase the mass of the disc over time (4\% after 400 orbits), also 
limiting the effective duration of these simulations.

Although gap depths for 
the \verb+2D30+, \verb+3D30I+, \verb+3D30A+, and \verb+3D30A+-Outflow 
do not match predicted values, they are 
consistent with each other, displacing $\sim$10-13\% of initial material in the 
vicinity of the planet.  We continue to use the results from these simulations 
throughout this paper, but consider the implications of if gap from 
the low-mass planet had reached its predicted depth in \S\ref{e30Synth}.

\section{Synthetic Images}\label{synthImages}

\begin{figure*}
\includegraphics[width=\textwidth]{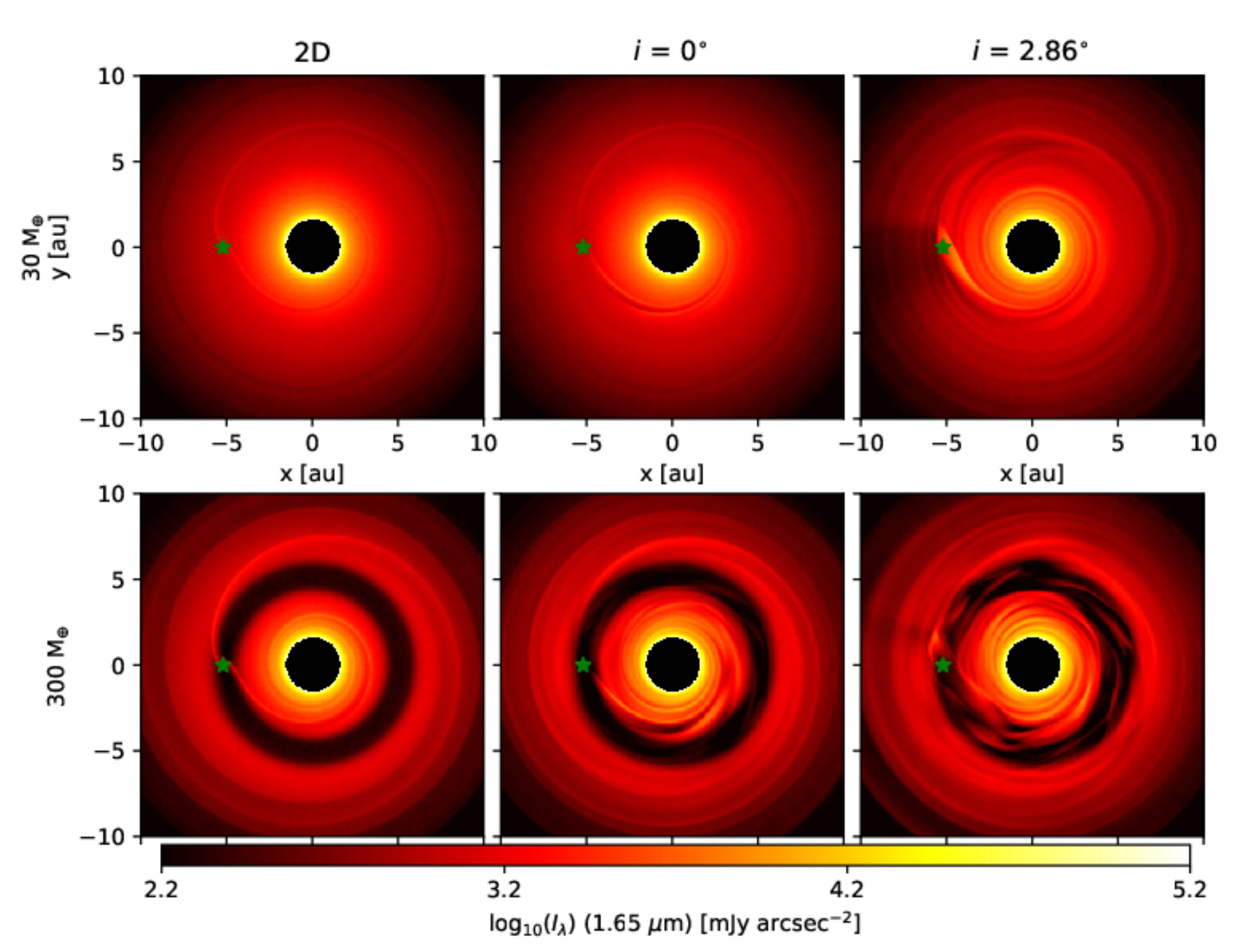}
\caption{Synthetic scattered light \emph{H-band} images of the 2D (Left), 3D non-inclined (Centre), 
and 3D inclined (Right) planetary-orbit simulations with 30 M$_{\earth}$ (Top) and
300 M$_{\earth}$ (Bottom) planets.  The actual location of the planet is indicated 
by a star symbol.}\label{scatMaps}
\end{figure*} 

We created synthetic images of how our model discs would be observed in
the \emph{H-band} (1.65 $\mu$m) using the
Monte Carlo radiative transfer code \verb+RADMC3D v0.41+ \citep{2012ascl.soft02015D}
(see Appendix \ref{RT_Setup} for details).

For the two-dimensional simulations we extrapolated a 3-dimensional density
profile from the resultant surface density values, $\Sigma(R,\phi)$, by assuming
surface density and density are related as
\begin{equation}
\rho(r,\theta, \phi)  =  \frac{\Sigma(R,\phi)}{\sqrt{2\pi} H} \
\exp\left(\frac{GM}{c_s^2}\left[\frac{1}{r} - \
\frac{1}{R}\right]\right),\label{twoTo3Rho}
\end{equation}
using the same 3D domain and grid, along with the $H$ and $c_s^2$ profiles,
described in \S\ref{initConds}.  We also
significantly decreased the density in the $10\times 10$ set of grid cells 
surrounding the location of 
the planet before making the 3-dimensional extrapolation to minimize  
the shadowing effect from having a physically unrealistic column 
of high-density material. 

\subsection{30 M$_{\earth}$ Planet Images}\label{e30Synth}

\begin{figure}
\includegraphics[width=\columnwidth]{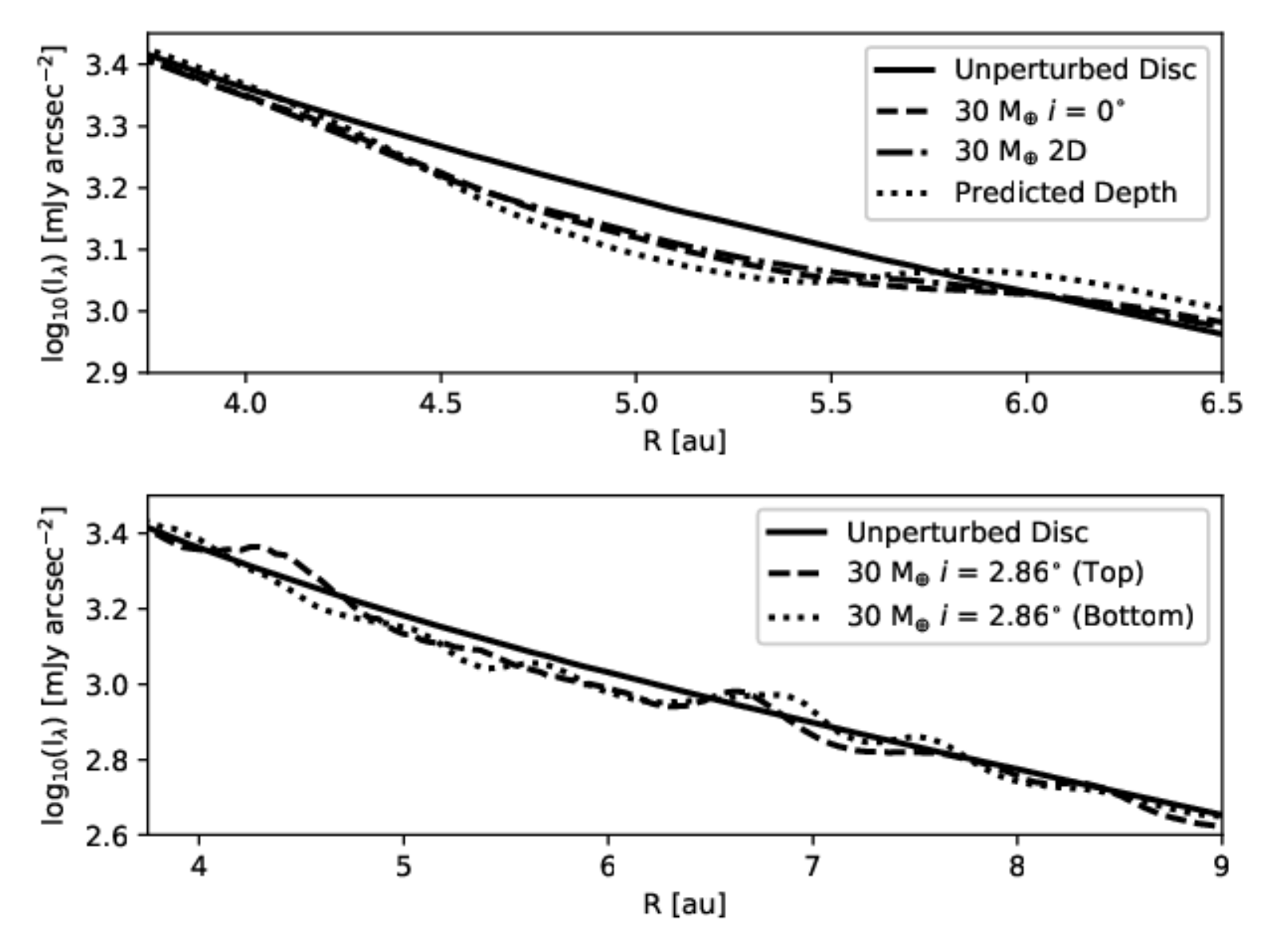}
\caption{Azimuthally averaged \emph{H-band} ($\log_{10}(I_{\lambda})$ $\lambda=1.65$ $\mu$m) 
scattered light 
from the 30 M$_{\earth}$ planet simulations.
{\bf Top:} Intensity dip at the location of the planet (5.2 au) for 3D30A model 
(Dashed, $--$), and the 2D30 model (Dash-Dotted, $-\cdot-$), 
compared to the intensity of an unperturbed disc (solid, $-$). 
Also included are results from simulation
in which the surface density gap reached the predicted depth from Eqn.(\ref{preDepth})
(doted, $\cdot\cdot$).  {\bf Bottom:} Intensity plot for the 3D30I model, extended out to disc radius of 9 au
to show global disturbances of the disc.  Dashed ($--$) line shows the averaged intensity
of the disc as viewed from the top, while the Dotted ($\cdot\cdot$) line is viewed from the bottom, 
compared to the intensity of an unperturbed disc (Solid, $-$).  Intensity features 
of the inclined planet show no obvious indication of a planet located at 5.2 au.}\label{scatAvgE30}
\end{figure}

In Figure \ref{scatMaps} (Top) we show the face-on \emph{H-band} synthetic images
produced from the results of the 30 M$_{\earth}$ planet HD simulations,
and plot the azimuthally averaged intensity profiles in Figure \ref{scatAvgE30}.  
Images from the \verb+2D30+ and \verb+3D30A+ models bear close resemblance to each
other, and to their respective surface density maps (Figure \ref{sdMaps}), 
including signs of their primary spiral density waves and secondary arms.
Unlike the surface density maps, there is no feature in the scattered light 
images which immediately indicate the displacement of material from the planet
(i.e., a dark annular region), 
however the effect of the surface density gap can be seen 
in Figure \ref{scatAvgE30} (Top), 
which shows a small decrease in average intensity near the vicinity of the planet.  
This intensity is 
87-88\% of the profile of the unperturbed disc, which corresponds with the depth of the 
surface density gap.  
This intensity dip seen in both the \verb+2D30+ and \verb+3D30A+ models suggests
that \emph{H-band} observation have the potential to
infer the presence of these lower-mass planets, as
such small deviations are found in VLT/SPHERE observations \citep[e.g.,][]{2018MNRAS.474.5105B}.  
Aside from this, the synthetic image of the \verb+3D30A+ model shows a distinct gap 
between the primary and secondary arms, making the existence of 
the secondary arm much more apparent. 

While the synthetic image of the 
\verb+2D30+ model does not reveal any more information than its corresponding
surface density map, and the \verb+3D30A+ image very closely resembles 
its own surface density map, the image produced from the
inclined 30 M$_{\earth}$ planet model (\verb+3D30I+) is significantly 
different.  In this image the primary arms are severely 
disrupted and the inclined orbit produces that appears around the location 
of the planet, which casts a shadow with an azimuthal arc of 30$^{\circ}$.  

In the case of 30 M$_{\earth}$ planet with an inclined orbit, there is also 
no obvious indication of the dark annular region observed in the images
from the non-inclined models.  This lack of intensity drop at the radial 
location of the planet is seen more clearly in Figure \ref{scatAvgE30} (Bottom),
which extends the average intensity profile out to 9 au.  
From this plot it is apparent that the intensity 
drops caused by planets with non-inclined orbits can be distinguished 
from the background, but disruptions from the inclined planet create 
several hills and valleys in this profile which are no where near the 
location of any planet.  

As mentioned in \S\ref{e30HD}, the maximum depth of the surface density gap
caused by a 30 M$_{\earth}$ planet is not quite as deep as predictions would indicate.
In running preliminary simulations used to test the models in this paper, 
we found that when running 2D simulations involving low-mass planets without the
increased gravitational softening parameter discussed in \S\ref{simSetup},
the depth of the gap did agree with predicted values ( $\Sigma\sub{min}/\Sigma_i = 0.80$ )
after only 200 orbits.  By extrapolating the results from this simulation
into a 3D distribution, 
then applying the radiative transfer process described in Appendix \ref{RT_Setup},
we can compare average intensity features of the predicted results to those from 
our models.  Results from this analysis are included in Figure \ref{scatAvgE30} (Top), 
and reveal an intensity that is approximately 82\% of the unperturbed profile.
This also corresponds fairly well to the depth of the surface density gap, and does 
demonstrate that small changes in surface density could have an effect on scattered 
light observations for ideal situations such as a single-planet system perfectly 
aligned with the disc.  Figure \ref{scatAvgE30} (Bottom) shows non-ideal situations
such as mis-aligned planetary orbits, do not produce a similar relation between
surface density and scattered light features.

\subsection{300 M$_{\earth}$ Planet Images}

\begin{figure}
\includegraphics[width=\columnwidth]{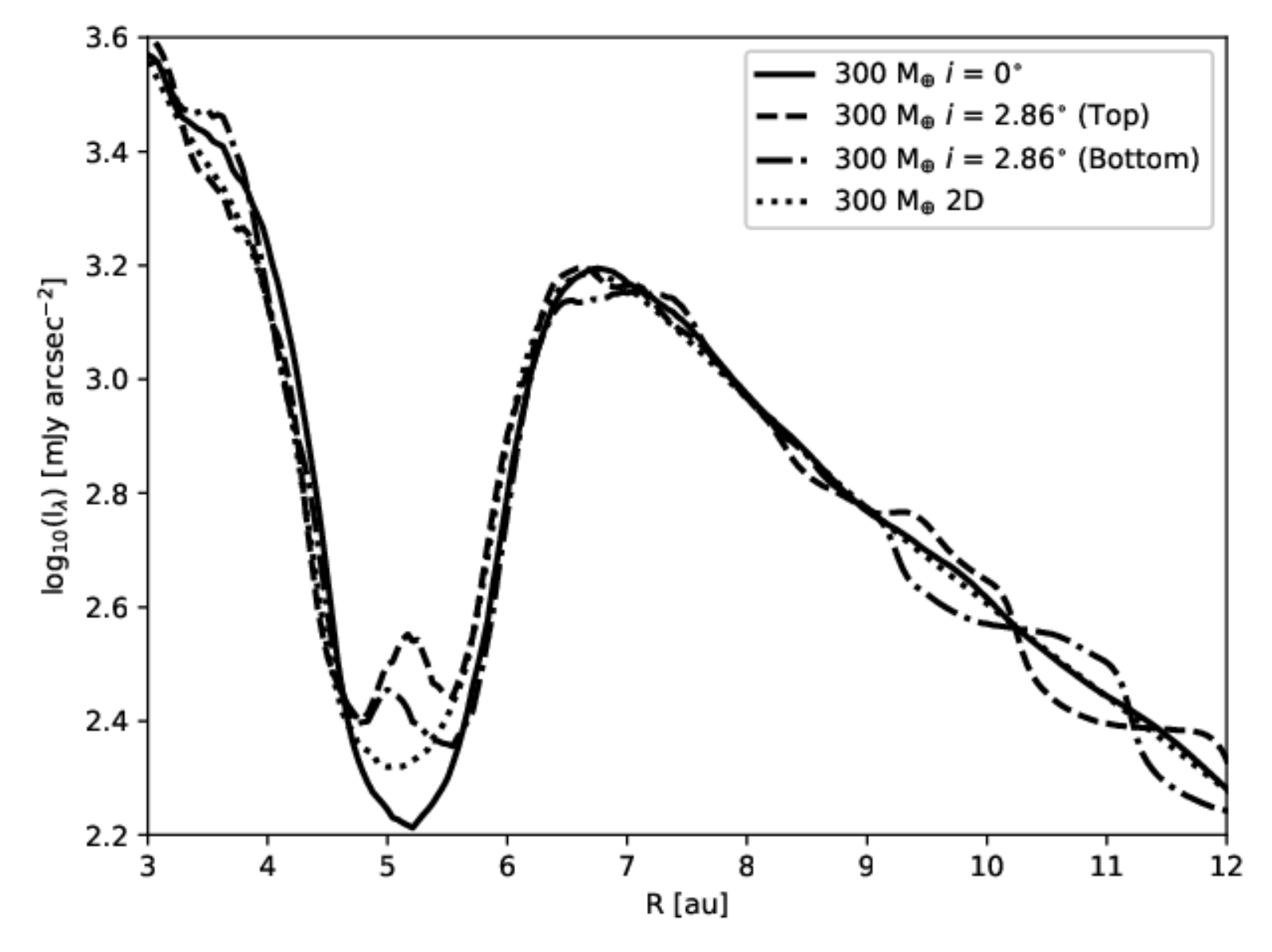}
\caption{Azimuthally averaged \emph{H-band}  ($\log_{10}(I_{\lambda})$ $\lambda=1.65$ $\micron$) 
scattered light intensity 
for the 300 M$_{\earth}$ planet.  Solid ($-$) line is intensity values from model 3D300A, 
while the Dashed ($--$) and Dash-Dotted ($-\cdot-$) lines are intensity values from
model 3D300I as view from the top and bottom of the disc, respectively.  Model 2D300 is 
represented by the Dotted ($\cdot\cdot$) line.  In all cases there is a clear and obvious dip 
at the location of the planet (5.2 au), however the maximum depth of the dip varies.  
The inclined planet also produces waves at the outer radius, where intensity peaks from
one point of view seemingly correspond to intensity dips from the other.}\label{scatAvgJup}
\end{figure}

Figure \ref{scatMaps} also shows the synthetic scattered light images 
from the 300 M$_{\earth}$ planet models, with their 
azimuthally averaged intensity plots in Figure \ref{scatAvgJup}.  As with
the 30 M$_{\earth}$ planet simulations, images produced from the \verb+2D300+ 
and \verb+3D300A+ models have similar characteristics with each other, and their
surface density maps (Figure \ref{sdMaps}).  The dark annular 
region at the planet's orbit is clearly apparent in both cases, 
along with the presence of primary, secondary, and tertiary arms.  
However, while the synthetic image from
the \verb+2D300+ model does not reveal any additional features compared to its 
surface density map, the features seen in the \verb+3D300A+ model image are slightly 
different; with signs of material remaining in the gap, as well as 
the emergence of a dark, spiral-arm-like feature 
near $(x,y) = (0,-4)$ au, between the primary and secondary arms that
makes the existence of the secondary arm much more apparent. 

Despite the similarities seen in their surface density features (Figures \ref{sdMaps} \& \ref{sdAvg}),
when their intensities are
azimuthally averaged over the entire disc, as shown in Figure \ref{scatAvgJup}, 
it seems that the slight differences between the surface density gap
depth from models \verb+2D300+ and \verb+3D300A+ (Table \ref{models})
produce obvious differences in intensity profiles.
When compared to the 
inclined model \verb+3D300I+, the dark annular region of the 
synthetic image in Figure \ref{scatMaps} shows a significant amount of turbulent
material within this region that is less apparent than in the other two models.
The spiral arms are also highly disrupted, 
which makes it more difficult to distinguish them from each other, or properly identify them.   
There is also a compact shadow-producing feature 
near $(x,y) = (-5.6,1.6)$ au.  The location of this feature is especially noteworthy
as its size and morphology could possibly be seen as evidence of a 
directly imaged, newly forming planet in an actual observation, however the planet in this 
simulation is actually located at $(x,y) = (-5.2,0)$ at the disc's midplane. 
The depth of the dip in the averaged intensity profile at the radial location of the 
planet for model \verb+3D300I+ is also not as deep as the other high-mass planet simulations, 
and the intensity shows hills and valleys at larger radii nowhere near the planet, 
similar to the inclined lower-mass planet simulation. 

\section{Discussion}\label{discussion}

\subsection{2D vs. 3D Simulations}

We used \verb+RADMC3D+ to calculate the 
height of the surface where, from the aspect of an observer, the optical depth
is
$\tau_s = 2/3$ for a wavelength $\lambda = 1.65$ $\mu$m
 ( i.e., the observed surface).
Applying this to our unperturbed density profile, we found that
this surface is 3.3 thermal scale-heights above the 
midplane at $R = R_0$.  

As the two-dimensional simulations are simply surface density
evolutions of disc-planet interactions, it is not surprising 
that producing synthetic scattered light images using density 
structures that have been vertically extrapolated from 
2D structures are essentially replications of these
surface density features.  This should be expected regardless
of the height of the surface.
The gas dynamics of the 3D simulations, however, are not 
necessarily vertically invariant.

\begin{figure}
\includegraphics[width=\columnwidth]{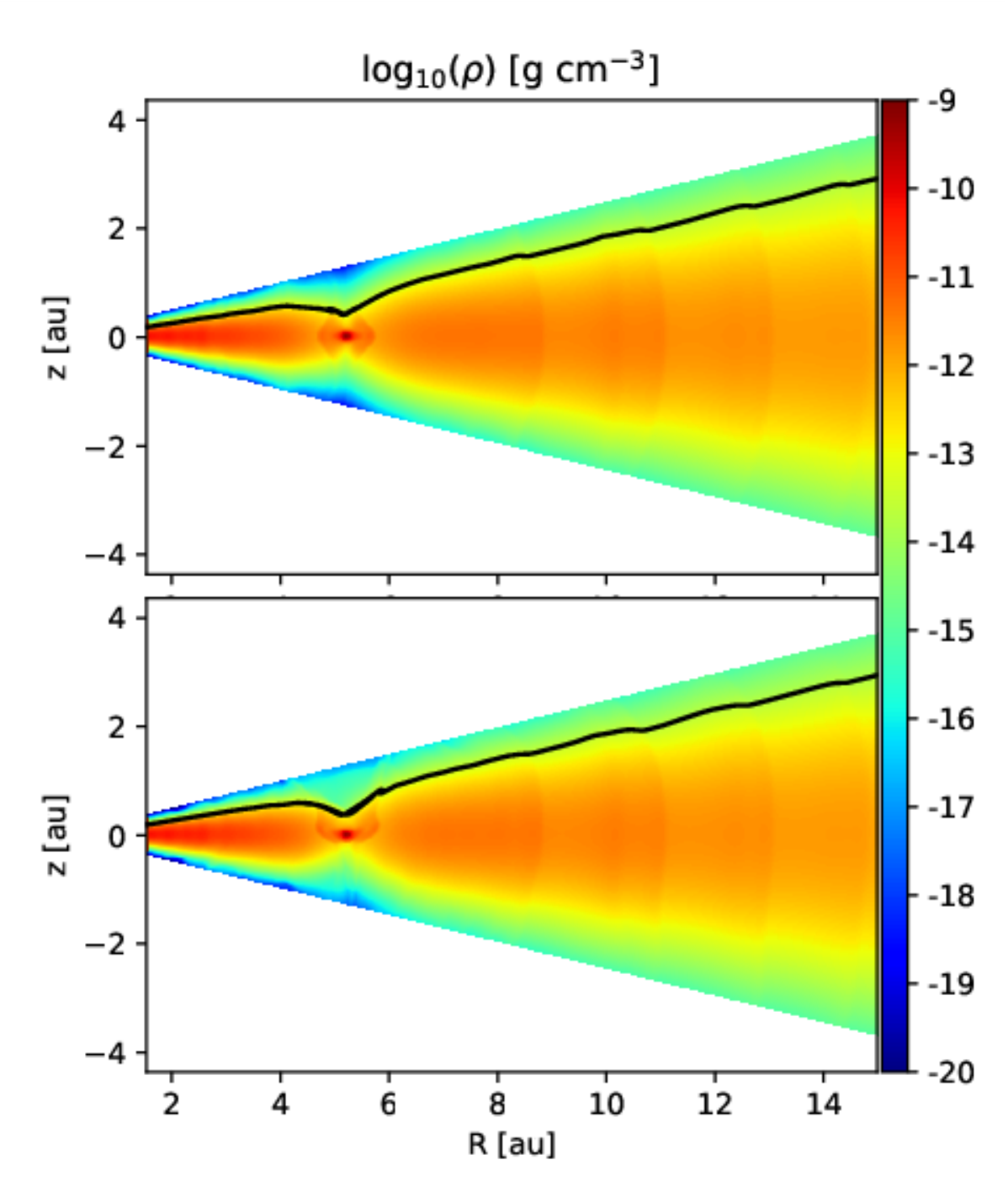}
\caption{Density map of vertical slice at the location of the 300 
M$_{\earth}$ planet for the $i=0^{\circ}$ simulation (Top) and the
$i=2.86^{\circ}$ simulation (Bottom).  The solid black line indicates the 
$\tau_s = 2/3$ surface of the disc for light of wavelength
$\lambda = 1.65$ $\mu$m.  }\label{sideView}
\end{figure}

Figure \ref{sideView} shows examples of vertical density maps of
the two 3D simulation results involving the 300 M$_{\earth}$ planet, 
as well as the profile of the observed surface superimposed onto
the map.  It should be noted that throughout this section we present 
lengths in au, with intensity and density values in cgs units, 
using the conversions described in Appendix \ref{RT_Setup}.
With this observed surface being so high above the 
disc midplane, it is also not surprising the surface density maps 
of the 3D simulations 
in Figure \ref{sdMaps}, essentially midplane features, are not 
exactly the same as their corresponding synthetic scattered light 
images.  
From this it would seem that \emph{H-band} images, synthetic and actual, 
reflect the disc's upper-atmospheric conditions.
To help confirm this, Figure \ref{topSlices} shows density slices 
at $\theta = 80.7^{\circ}$ (the height of the scatter surface
of an unperturbed disc 
at $R=R_0$) from the results of  
our four 3D models.  The features seen in these 
upper-atmosphere
density maps much more closely resemble those in the scattered-light images
of Figure \ref{scatMaps} than the surface density maps in Figure \ref{sdMaps}.
Although not visible in the 2D simulation, these features include 
wisps of material can be clearly seen in the annular ring created 
by the 300 M$_{\earth}$ planet in the 3D simulations.  This demonstrates 
the possibility that despite a planet almost completely clearing out
disc material at the midplane, gas remnants could still be 
present in the gap at the disc's upper-layers; a feature models
from 2D simulations would miss.

\begin{figure}
\includegraphics[width=\columnwidth]{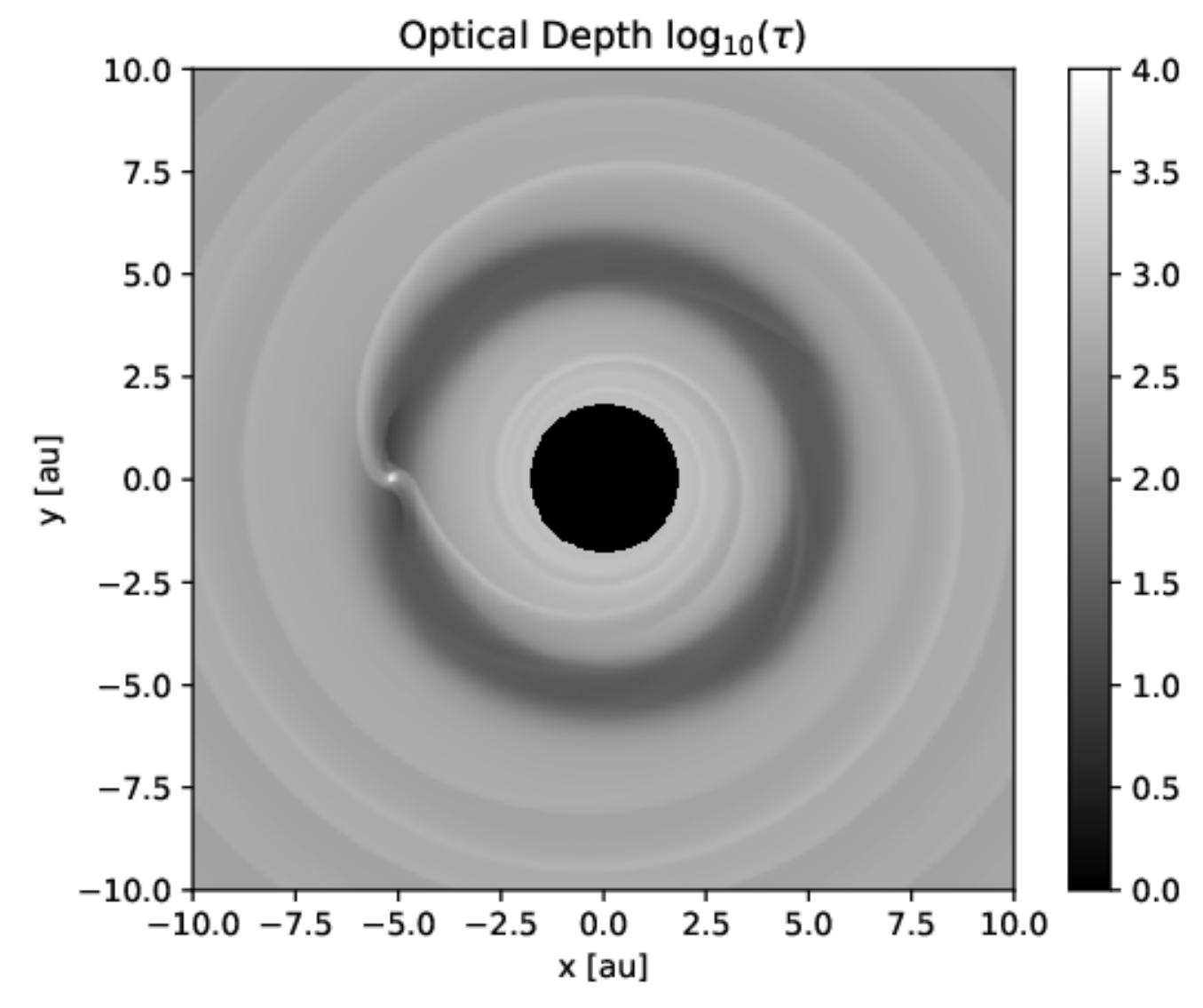}
\caption{$\log_{10}$ map of optical depth for light of wavelength $\lambda = 1.65\micron$ 
at the midplane of simulation 3D300A.  
In general, the optical depth at $R = R_0$ (the center of the gap)
is $\tau\approx 20$.}\label{tauMap}
\end{figure}

\begin{figure}
\includegraphics[width=\columnwidth]{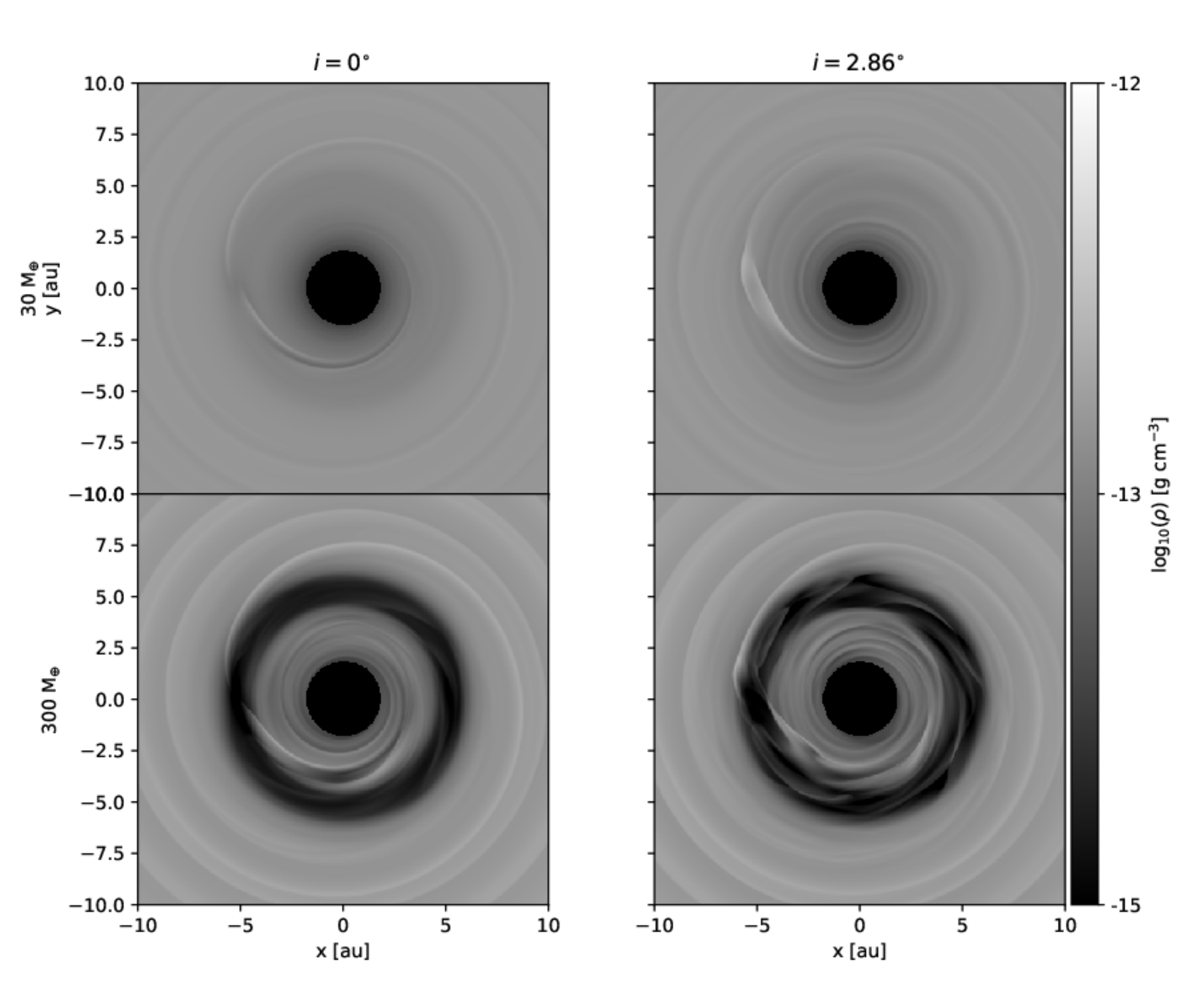}
\caption{Density maps of the 3D simulations cut along the polar angle 
of $\theta = 80.7^{\circ}$, corresponding to the angle of the 
$\tau_s = 2/3$ surface of an unperturbed disc at $R=R_0$.
Results shown are for planets with non-inclined orbits (Left), and
inclined (Right) for simulations involving a 30 M$_{\earth}$ planet
(Top) and a 300 M$_{\earth}$ (Bottom).
These upper-atmosphere 
density maps resemble the synthetic images seen in Figure \ref{scatMaps}
much more than those of the surface density maps in Figure \ref{sdMaps}.}
\label{topSlices}
\end{figure}

To demonstrate the difficulty of near-IR observations to 
see into the midplane, we approximate the optical depth 
from an observer to the midplane via,
\begin{equation}
\tau_{\lambda} = \int_{\theta_0}^{\pi/2} \rho_d \kappa_{\lambda} r~d\theta,
\end{equation} 
where $\theta_0$ is the upper polar boundary of the disc, $\rho_d$ is the
dust density, 
and $\kappa_{\lambda}=1320$ cm$^{2}$ g$^{-1}$ is total dust opacity at a wavelength 
$\lambda = 1.65\micron$ (from Figure \ref{opacities}).  Figure \ref{tauMap} shows this applied to
the \verb+3D300A+ simulation.  Certain areas near the planet in the 
$\pm$y-direction have an optical depth as small as $\tau=10$, but at the 
orbital radius of the planet the optical depth is $\tau\approx 20$, even when
94\% of the initial material has been removed.  Directly
observing midplane features in these discs at near-IR wavelengths, 
or directly observing a planet within the disc, would not be feasible.

That these scattered light images are looking at the upper
layers of protoplanetary discs raises the question of whether synthetic
images produced from 2D simulations can reliably model the observable 
near-IR features of these discs.  
Works such as \cite{2014A&A...572L...2R}  and \cite{2015ApJ...813...88Z} 
argue that in order to realistically model these discs, 
and their observable properties, fully 3D models are required
for reasons such as that 2D simulations can ``overestimate the 
detectability of planet-induced gaps'' and spiral arms 
are much more prominent in synthetic images produced from
3-dimensional simulations.
On the other hand, \cite{2016ApJ...832..105F} compare 
surface density profiles of 
2D and 3D hydrodynamic simulations to show they are similar, and 
\cite{2017ApJ...835..146D} compare 
synthetic \emph{H-band} images of an example
Saturn-mass planet produced from 2D and 3D hydrodynamic simulations 
to justify the use of 2D simulations for modeling observable 
disc features.  Our results seem to be in agreement 
with the idea that under ideal conditions, 3D density extrapolations 
from 2D simulations are sufficient to create models of 
scattered light images, as the uncertainty of planet 
mass estimates from disc observations is typically quite large
\cite[e.g.,][]{2017A&A...600A..72F,2018A&A...617L...2M}.  For interactions in which 
the planet's effects on the disc are not vertically 
invariant, however, models from fully three-dimensional 
simulations are required.
 
\subsection{Inclined Orbits}

Figures \ref{sdMaps} and \ref{sdAvg} show that in the cases 
of the 3D hydrodynamic simulations, slight planet inclinations 
seem to have no effect on the features seen in the surface density
maps, or the average surface density profile, for both the high-
and low-mass planets.  However, Figures \ref{scatMaps}, 
\ref{scatAvgE30}, \ref{scatAvgJup}, and \ref{topSlices} all show 
that an inclined planet does have a significant impact on the gas
dynamics of the discs at the height of the scatter surface, 
as well as a significant impact on the scattered light features
and intensity of near-IR observations.  The ability for an inclined
planet to alter the gas dynamics of the upper-atmosphere is 
expected considering the added vertical component of the planet's 
motion.  
As long as small ($\sim$$\micron$) dust grains are able to remain 
coupled to the turbulent gas at the surface, this significant
effect on the near-IR observable features is also expected.  
We explore how reducing the amount of dust in the upper atmosphere 
would affect our observations in Appendix \ref{lowDust}.    

Figures \ref{scatAvgE30} and \ref{scatAvgJup}
also show that there are large radial variations in average intensity 
throughout the disc for the inclined-planet simulations.  
As the depth of some of these intensity variations in
the \verb+3D300I+ model are greater than the intensity gap 
produced in \verb+3D30A+, 
in actual observations these kinds of variations could be falsely perceived as
signatures of multiple planets, or even as secondary gaps in the disc due to 
shock waves caused by a single massive planet \citep[e.g.,][]{2017ApJ...850..201B}.
Models attempting to interpret such observations using the assumption that the planet
is co-orbital with the disc would obtain incorrect results.  
Likewise, these large radial variations in intensity are not necessarily 
evidence for an inclined planet.  As seen in Figure \ref{scatMaps} (Bottom-Right), 
the variations in average intensity are likely caused by the disruption of 
the spiral density arms in the disc's upper-atmosphere, which could be the result
of any regular disruption.   

The density maps of Figure \ref{topSlices} and the 
the synthetic scattered light images of Figure \ref{scatMaps}
show that not only do scattered light images show the 
upper-atmospheric features of the disc, but that some of these 
features cause shadows which could affect the temperature 
and overall dynamics of the disc. One of these features 
is particularly noteworthy; a small bright area which resembles an overdense 
region, and causes a shadow, in the inclined 300 M$_{\earth}$ planet simulation
(Figure \ref{scatMaps}, Bottom-Right).
A similar feature, a bright point-like object, has been observed in the $L'$-band 
in the transitional disc HD169142 which was considered to be a 
potential 28-32 M\sub{Jup} companion object, however the absence of 
this bright point source in $H$-band or $K_s$-band makes it more likely
that is is a disc feature and not a planet
\citep{2014ApJ...792L..22B, 2014ApJ...792L..23R}.  Our simulations have 
produced a bright point-like feature which could be interpreted as 
a newly forming planet or brown dwarf companion object, however 
we know for a fact that there is no planet at the location of 
this feature.  

The cause of this bright area in \verb+3D300I+ is a high density 
region of gas near the upper-boundary of the simulation that 
appears to be connected to the outer spiral density wave.
This could potentially be a result of the disc-planet exchange 
which produces spiral density waves is occurring at varying levels 
of the disc, creating a much more complicated dynamic.  
This amorphous high-density region is located between 4-5$H$ above (but not below) the disc 
midplane, but begins to disappear at around 3.5$H$ (Note this 
high-density area is not as prominent in Figure \ref{topSlices}, 
which is a density slice at 3.3$H$).  
While the exact cause of this feature unknown, the feature appears 
quite early in the simulation ($\sim$30 orbits), which gave us
the opportunity to run several short exploratory trials.  

\begin{enumerate}
\item Increased the cell numbers in the radial and azimuthal 
dimensions to provide uniform grid resolution at $R = R_0$.  
\item Radial boundary conditions were set to outflow.
\item Outflow conditions at the polar boundaries were slightly altered 
such that the density values in the ghost cells were linearly extrapolated 
from domain values, rather than simple zero-gradient.
\end{enumerate}

In all cases this high-density region persisted.  
We also decreased the vertical size of the simulation to
$\pm$4$H$ at $R = R_0$, as well as $\pm$3$H$ at $R=R_0$.
The density layers at the upper-boundaries of these smaller 
simulations matched very closely to those of the 
4$H$ and 3$H$ density slices from the full simulation.  The 
high-density region appears at the upper boundary of the 4$H$ 
simulation, but not at the upper boundary of the 3$H$ simulation, 
indicating that it is not simply due to an interaction between the gas and 
the boundary.    

Analysis of radial, polar, and azimuthal velocities show that the gas
dynamics at these layers of a disc with an inclined planet are far more
complex than with an aligned planet.  It is apparent that the high density
region arises from this complex motion of the gas. 
By investigating with different boundaries this feature still remained. 
However we note due to the limitation of the model (e.g., simulating
these complex upper-atmospheric dynamics at these
scale heights using the vertically isothermal approximation) 
the physical nature of this feature should be taken with care.

\begin{figure}
\includegraphics[width=\columnwidth]{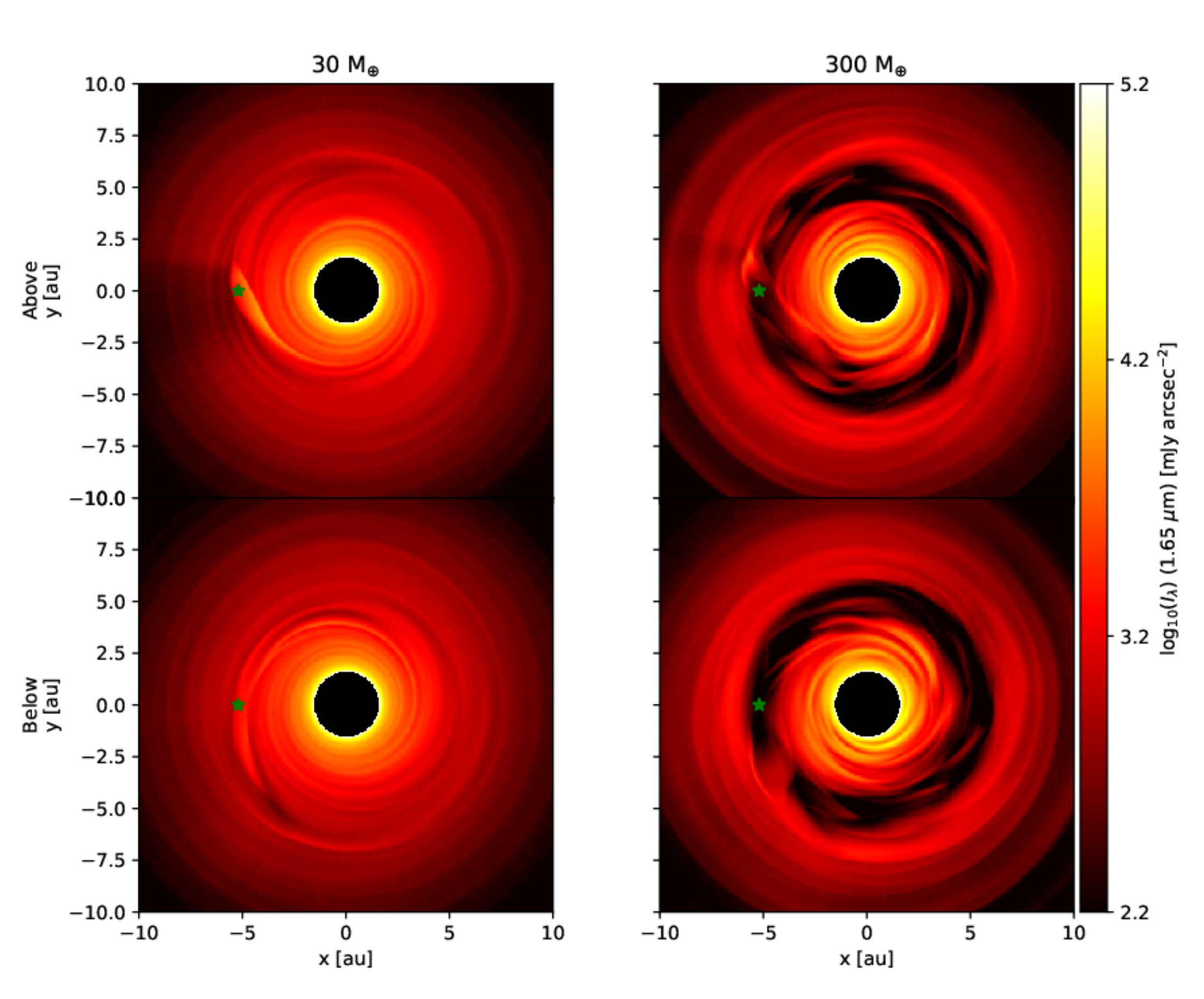}
\caption{ Synthetic \emph{H-band} (1.65 $\mu$m) scattered light maps involving the 
inclined 30 M$_{\earth}$ planet (Left) and the inclined 300 M$_{\earth}$ 
planet (Right) viewed from above (Top) and below (Bottom), 
with a star symbol indicating the location of the planet.  The images are 
mapped such that from their respective
points of view, the planet is rotating counter-clockwise as viewed from above, 
while it is rotating clockwise as viewed from below.  }\label{rotated}
\end{figure}

The shadow-causing feature seen in the synthetic image 
of the \verb+3D30I+ model is also caused by a high-density region
in the disc's upper-atmosphere (Figure \ref{topSlices} Top-Right).  
This feature also follows the spiral density waves, and its
origin is likely related to them and the complications that arise 
with these arms being produced at different heights in the disc.  
This dense region is extended, and in turn causes an extended 
shadow over a large part of the disc.  The implications of this 
extend beyond the observable features discussed in this work, 
as the cooling of such a large region could have a profound 
effect on the evolution of the disc.  To fully explore the 
consequences of shadowing on this scale, 
radiation hydrodynamic, or even radiation magnetohydrodynamic, 
simulations are required \citep{2013A&A...560A..43F}.   

Another implication for observations is the asymmetry of the 
scattered light surfaces due to the inclined orbit of the planet.
While azimuthally averaged intensity profiles do depend 
slightly on the inclination of the disc with respect to the 
observer \citep{2017ApJ...835..146D} if the orbit of a planet 
is aligned with the midplane, scattered light observations are 
essentially the same as viewed from the top of the disc as those 
viewed from the bottom.  This is not the case 
when the the disc is asymmetric across the midplane, as demonstrated
in Figure \ref{rotated}, which shows how the discs with inclined 
planets would be observed in scattered light as viewed from above 
and below.  Despite that both points of view would 
be considered face-on observations, they show striking differences 
between each other; most notably the lack of a shadow cast by 
the disrupted spiral arm as viewed from the bottom 
in the lower-mass planet simulations. 
These different points of view also produce differences in average
intensity profiles, as shown in Figures \ref{scatAvgE30} \& \ref{scatAvgJup}.  
Viewed from below, there is no obvious indication that there is a planet 
located at 5.2 au for the \verb+3D30I+ model, and the intensity profile 
produces variations in intensity similar to those seen as viewed from the 
top.  In the case of the \verb+3D300I+ model, there is a clear indication
of a planet at 5.2 au, which produces a dip in intensity slightly greater 
than its top-view counterpart.  The view from the bottom also produces an interesting 
feature in that the peaks and valleys in intensity radially outwards from the 
planet are directly opposite to those as viewed from the top.  As this 
is not seen in the lower-mass planet situation, further study is required
to understand under what circumstances this occurs, or if it is simply 
an isolated incident.  Comparing these results to follow up simulations 
in which the sign of Eqn. \ref{z_t} is changed would also be a worth-while 
endeavor.

\begin{figure}
\includegraphics[width=\columnwidth]{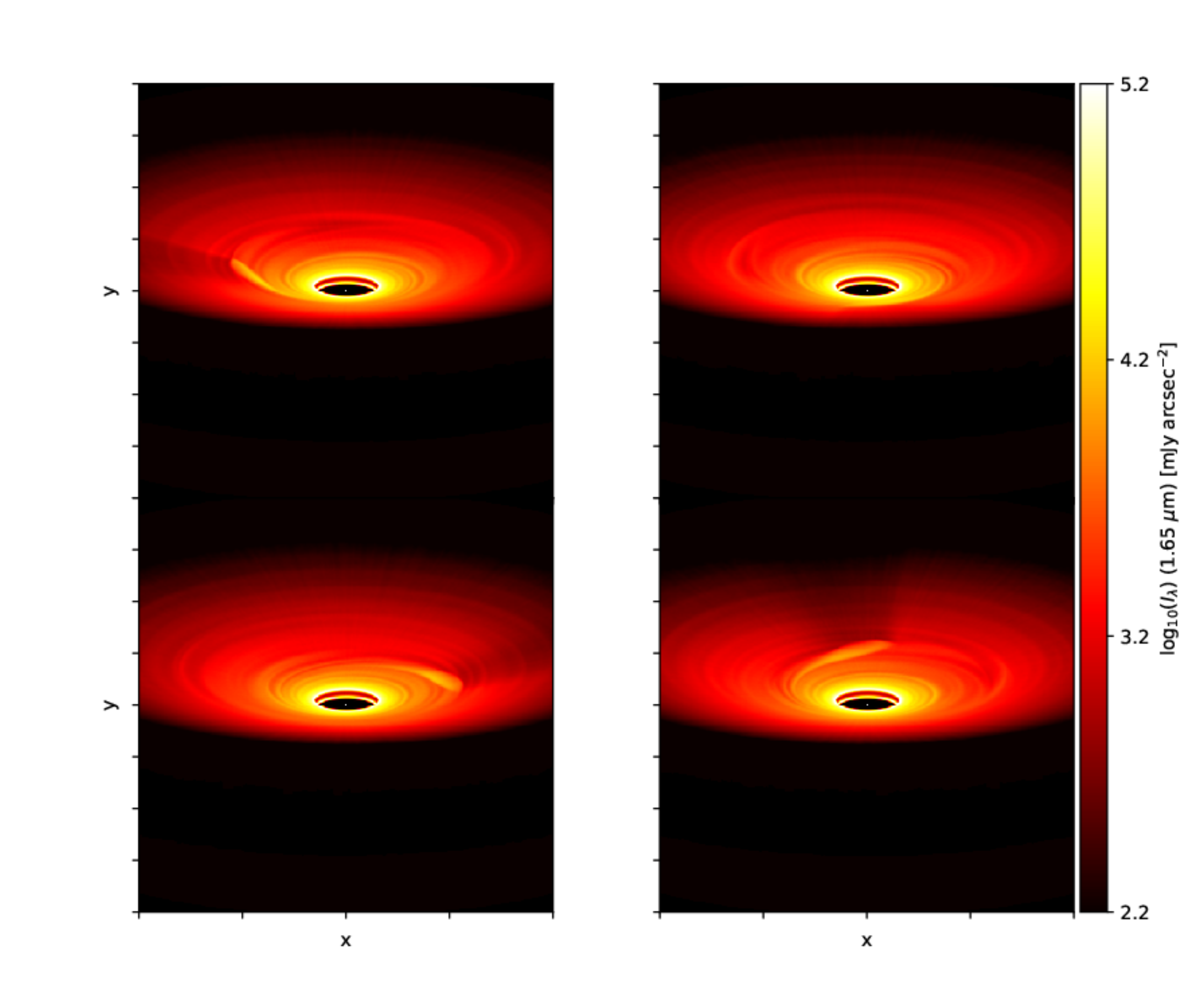}
\caption{Synthetic \emph{H-band} (1.65 $\micron$) scattered light maps from the 
inclined 30 M$_{\earth}$ planet simulation (3D30I), observed with 
the disc inclined at $i_d = 70^{\circ}$ (Top Left).  Other images represent how the 
disc would be observed if it were rotated azimuthally anti-clockwise by 
$90^{\circ}$ (Top Right), $180^{\circ}$ (Bottom Left), and $270^{\circ}$ (Bottom
Right).}\label{inclinedDisk}
\end{figure}

In spite of the shadow from features caused by inclined planets seen in 
Figure \ref{scatMaps} can not be seen when the disc is viewed from below,
Figure \ref{inclinedDisk} shows that the massive massive shadow from the 
\verb+3D30I+ model is 
visible even when the disc itself is inclined by $i_d = 70^{\circ}$ from face-on.
The shadow can also be seen at almost any rotational alignment of the disc,
except when the disc is orientated such that the planet is between the 
star and the observer.  It would seem that observations 
of discs with an embedded mis-aligned planet depend not only
on disc inclination and orientation, but also whether the disc is being viewed
from above or below.

\subsection{Applications to Observations}\label{observations}

\begin{figure}
\includegraphics[width=\columnwidth]{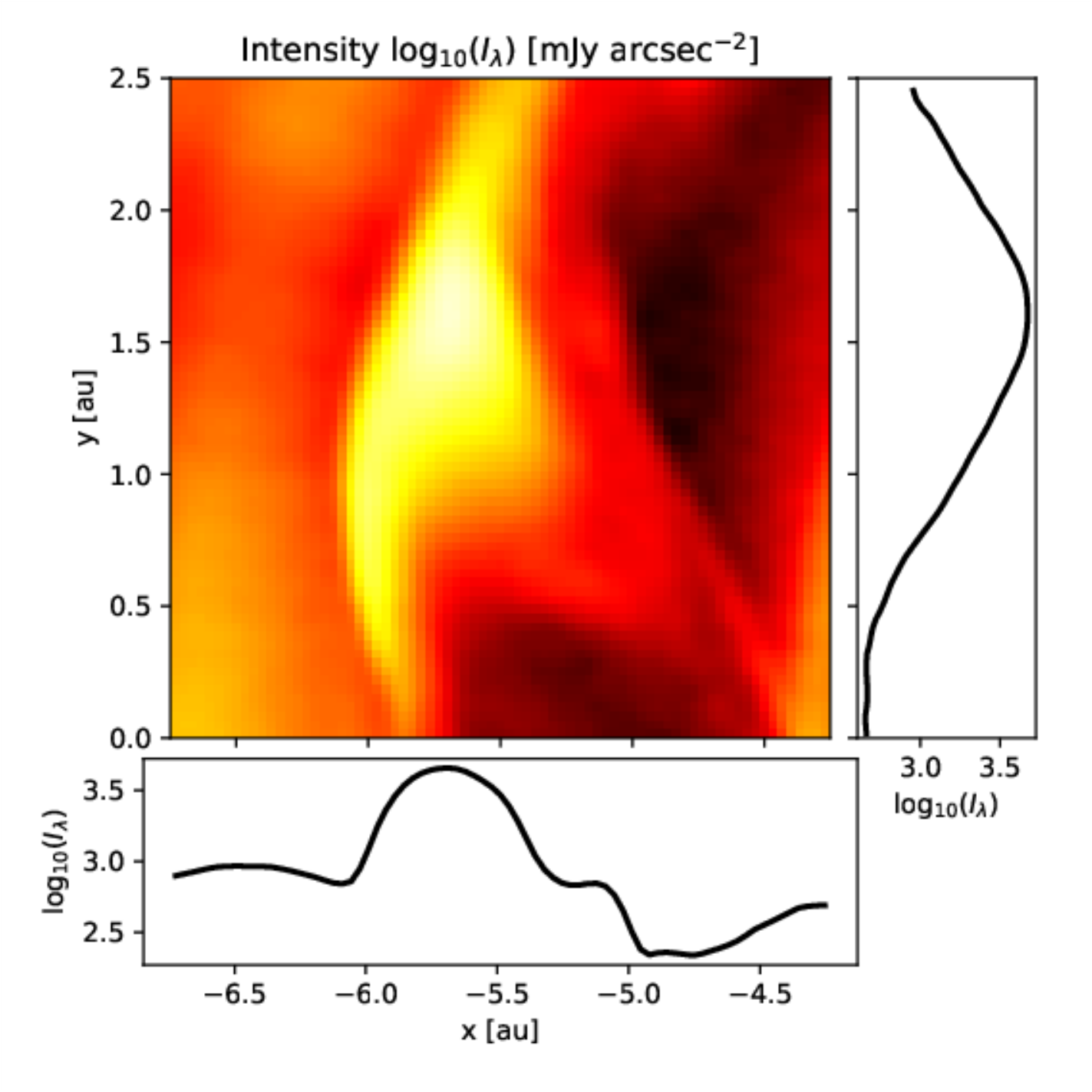}
\caption{Intensity ($\lambda = 1.65 \micron$) map focused on bright region seen in 
the synthetic scattered light image of model 
3D300I, along with a vertical intensity plot sliced at 
$x = -5.7$ au (Right) and a horizontal intensity plot 
sliced at $y = 1.5$ au.  The intensity of this bright region 
can be 20 times higher than the surrounding areas.}\label{brightContrast}
\end{figure}

The bright features in our synthetic \emph{H-band} images
of discs with an embedded inclined planet are clearly 
visible in contrast to their surrounding intensity levels.
This is demonstrated in Figure \ref{brightContrast}, which
shows an intensity map of a $2.5 \times 2.5$ au$^{2}$ area surrounding the 
small bright region seen near the planet in the \verb+3D300I+
model.  This map is accompanied by intensity slices at $x = -5.7$ au 
(Figure \ref{brightContrast} Right), and $y = 1.5$ au (Figure \ref{brightContrast} 
Bottom).  The intensity levels here can be up to twenty times higher than those in its immediate vicinity.
This is also true in the feature seen in \verb+3D30I+, which is 
also larger and produces a shadow on the disc radially outwards from
itself.  

To explore the feasibility of actually observing these features, 
we placed them face-on at a distance of 100 parsecs and convolved them 
using a resolution of 0.03$''$ to approximate the capability 
of \emph{H-band} observations from VLT/SPHERE.  
No features were resolvable at these scales, with the exception 
of the large shadow in the \verb+3D30I+ model.  This indicates 
that it is possible to detect extremely large-scale shadows 
without being able to identify the cause.   
It should be noted that \cite{2017ApJ...837..132V} has been able 
to resolve gaps at radii of less than 5.2 au using VLT/SPHERE
for TW Hydrae at distance of 54 parsecs using polarized light. 
Synthetic scattered light images simulating polarized light at these 
scales will be considered for future work.   
While other instruments used to view protoplanetary discs,
such as ALMA, have the resolution to observe a disc gap in TW Hydrae 
at 1 au \citep{2016ApJ...820L..40A}, 
these observations are at longer wavelengths which penetrate much deeper into the disc.

\begin{figure}
\includegraphics[width=\columnwidth]{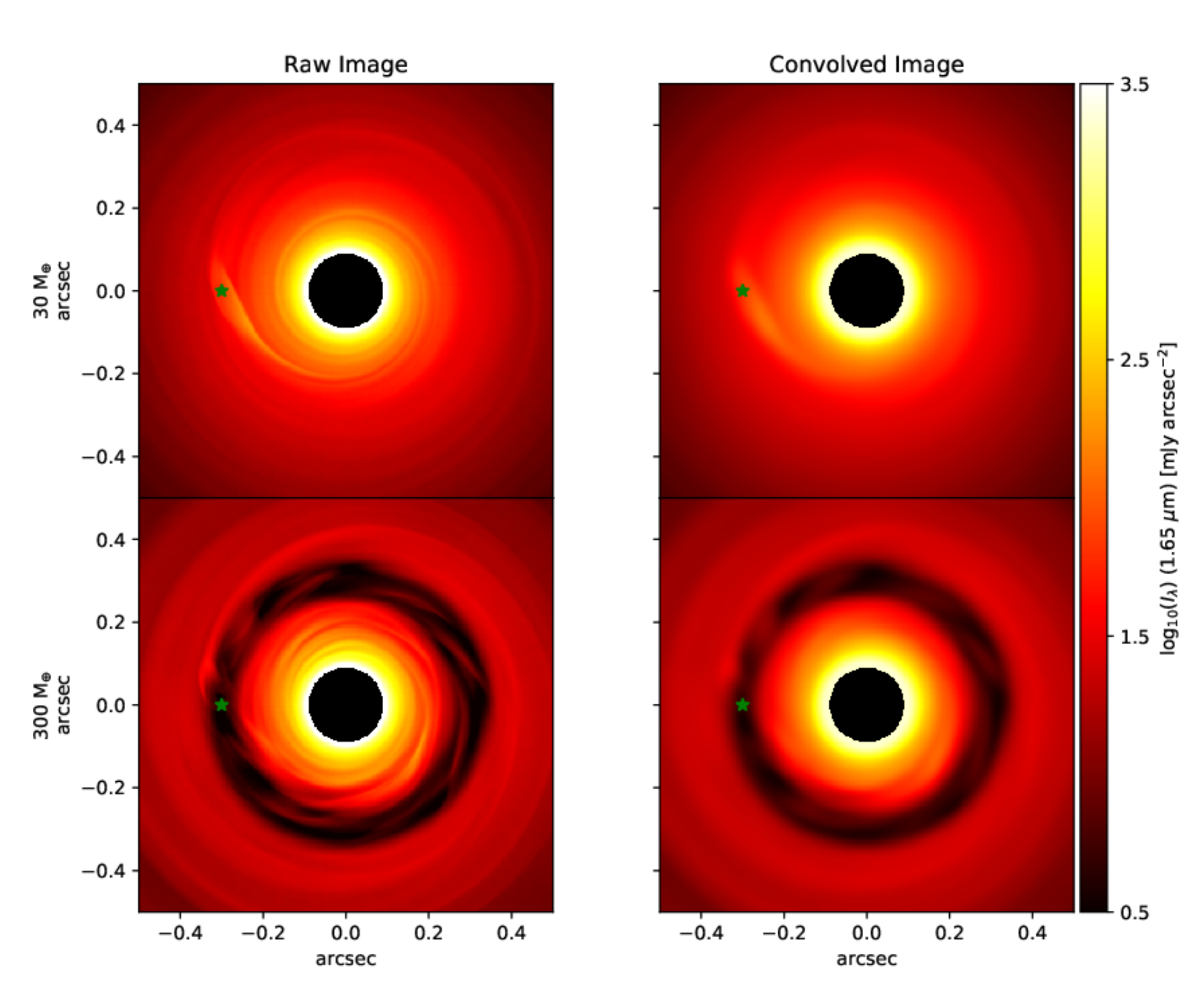}
\caption{Synthetic \emph{H-band} ($\lambda = 1.54\micron$) images of 
the inclined 30 M$_{\earth}$ (Top) and 300 M$_{\earth}$ (Bottom) planet 
simulations scaled to $R_0 = 30$ au.  Raw images are shown on the Left, 
while images convolved to resemble discs at 100 parsecs observed 
with a resolution of 0.03$''$ are shown on the Right.}\label{thirtyAU}
\end{figure}

\begin{figure}
\includegraphics[width=\columnwidth]{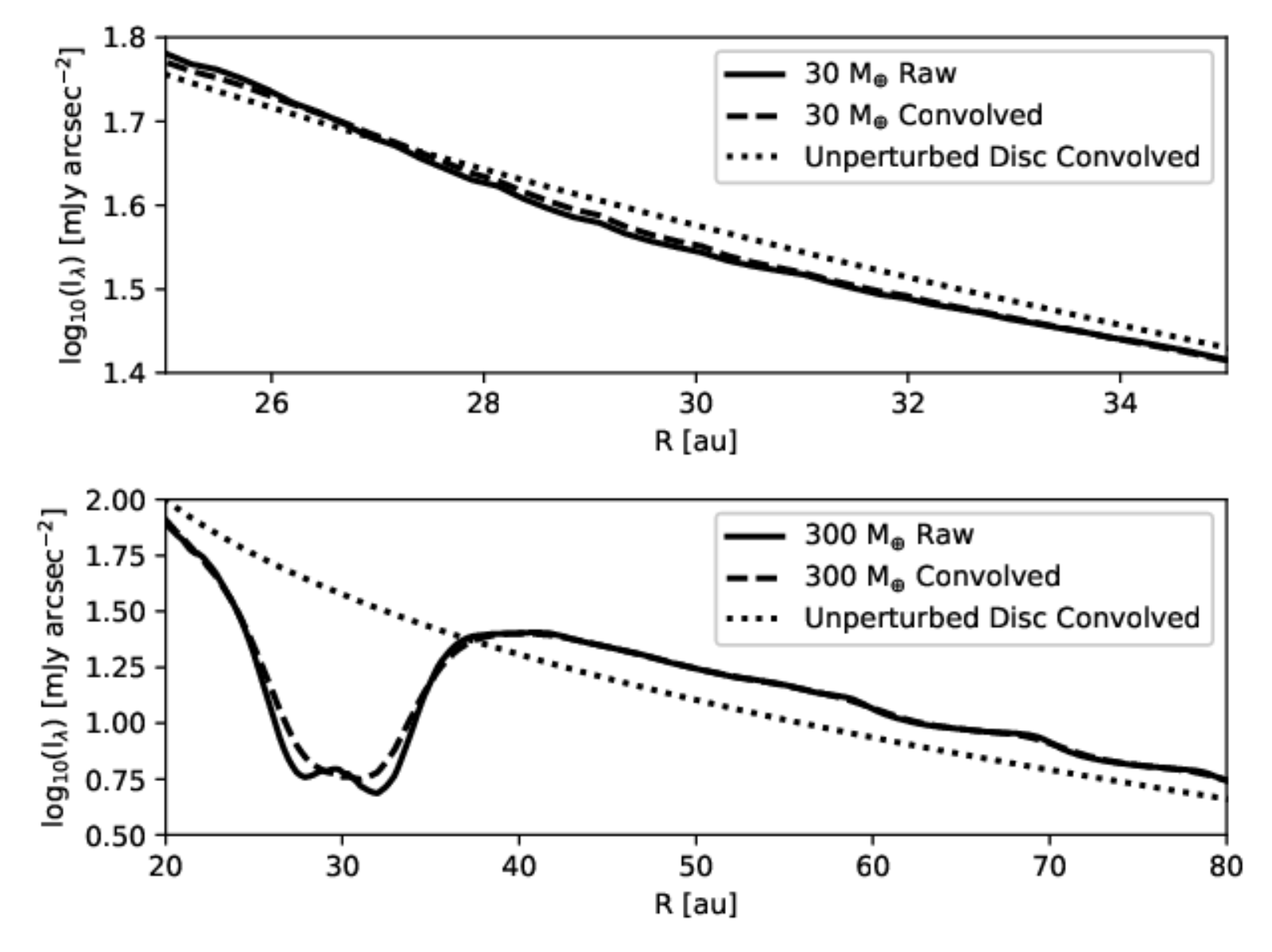}
\caption{Azimuthally averaged intensity values for images 
of discs involving the inclined 30 M$_{\earth}$ planet (Top) and
inclined 300 M$_{\earth}$ planet (Bottom) at 30 au.
The raw images (Solid, $-$) are compared to convolved 
images (Dashed, $--$), and to convolved images of 
unperturbed discs (Dotted, $\cdot\cdot$).}\label{avgConv}
\end{figure}

We also scaled the dust density from the \verb+3D30I+ and \verb+3D300I+
models and scaled them out to emulate our results for a planet orbiting
at $R_0 = 30$ au, with a surface density of $\Sigma_0 = 30$ g cm$^{-2}$
and an aspect ratio of $H/R = 0.07$ 
at $R = R_0$.  With these new dust density profiles, we again used \verb+RADMC3D+
to create temperature and synthetic \emph{H-band} images using the process 
described in Appendix \ref{RT_Setup}.  We then convolved these images
using the same distance and resolution described above (Figure \ref{thirtyAU}).
The feature caused by the 30 M$_{\earth}$ planet is still clearly visible, 
but no longer casts an obvious shadow.  It is also distinguishable in
the convolved image.  There is also no obvious dark angular region at the orbital 
radius of the planet, although when the disc intensity 
is azimuthally averaged, as in Figure \ref{avgConv}, the average intensity
at the planet is 96\% of an unperturbed disc.  

The small bright feature from the \verb+3D300I+ model is not as apparent as its 
5.2 au counterpart, however, and is essentially non-existent in the convolved image.  
This is similar to the lower dust-to-gas models discussed in Appendix \ref{lowDust}.
In this case the synthetic image also sees deeper into the disc, 
with a $\tau = 2/3$ surface this is at only $0.7H$ at $R= R_0$, which effectively
ignores aspects of the disc at the upper boundary.  The average
intensity is 15\% of an unperturbed value, and the dark annular region 
at the location of the planet can be clearly seen in our convolved intensity map.  
Perturbations 
at the outer radius are also still apparent, similar to those seen in Figure \ref{scatAvgJup}.
These dips can be as high as 10\%, which are comparable to the depth caused by the 30 M$_{\earth}$
planet so could be mistaken for secondary planets.  

Unfortunately there are severe limitations implementing radiative 
transfer techniques to this kind of scaling, namely in the difference 
of aspect ratios at 5.2 and 30 au.  With our polar bounds, an aspect 
ratio of $H/R = 0.07$ at 
$R_0 = 30$ au limits the vertical domain of the disc to $\pm 3.57 H$ at the
location of the planet. Evolving a full disc with more suitable initial 
density conditions could effect the dynamics in the upper-atmosphere, 
and including the dust distribution of the upper layers could raise the 
location of the observed surface.  Future work involving observational
features of protoplanetary discs caused by planets will focus on this region.

\subsection{Damped Orbital Inclination}\label{damping}

Although the difference in observable scattered light features 
between discs containing inclined and non-inclined planets is 
obviously apparent, 
other HD simulations have explored the evolution of inclined planets
\citep[e.g.,][]{2007A&A...473..329C,2011A&A...530A..41B} and have found that for 
low-inclined planets (i.e. $i \la H/R$ ), orbital inclination of a planet
decays exponentially, $di/dt \propto -i$.
By varying a planet's initial inclination
and mass, and allowing the planet to move throughout the simulation, 
\cite{2011A&A...530A..41B} determined that the damping time-scale was 
$\tau_d = 53$ orbits for a planet at 5.2 au in an isothermal disc 
with a thermal scale-height of $(H/R)_{R_0} = 0.05$, and $\tau_d = 27$ orbits
when radiative transfer effects where taken into account throughout the 
simulation.  To explore how damping inclination would affect scattered light
features, we re-run our 3D simulations of inclined planets with an artificial
damping of 
\begin{equation}
i = i_0e^{-(t-10)/\tau_d} \text{  for } t \gid 10 \text{ orbits},
\end{equation}
where $i_0 = 2.86^{\circ}$.  The inclusion of $(t-10)$ in the exponential is to 
account for the fact that we still grew the planet mass over the course of 10 
orbits and did not begin to damp the orbital inclination until the planet 
had reached full mass.

In this new case we run the simulations with $\tau_d = 53$ orbits
for the 30 M$_{\earth}$ and 300 M$_{\earth}$ planets for 200 orbits, and create synthetic 
images at snapshots of $t = 100$ and $t = 200$ orbits.  For comparison, we also create synthetic 
images at $t = 100$ and $t = 200$ orbits for all previous 3D simulations. 
In addition, despite the fact the simulations presented in this paper 
are vertically isothermal, we also run the damped inclination simulations using the 
time scale determined from the \cite{2011A&A...530A..41B} radiative transfer simulations 
of $\tau_d = 27$ orbits. However, we only run these simulations for 100 orbits.  

\begin{figure*}
\includegraphics[width=\textwidth]{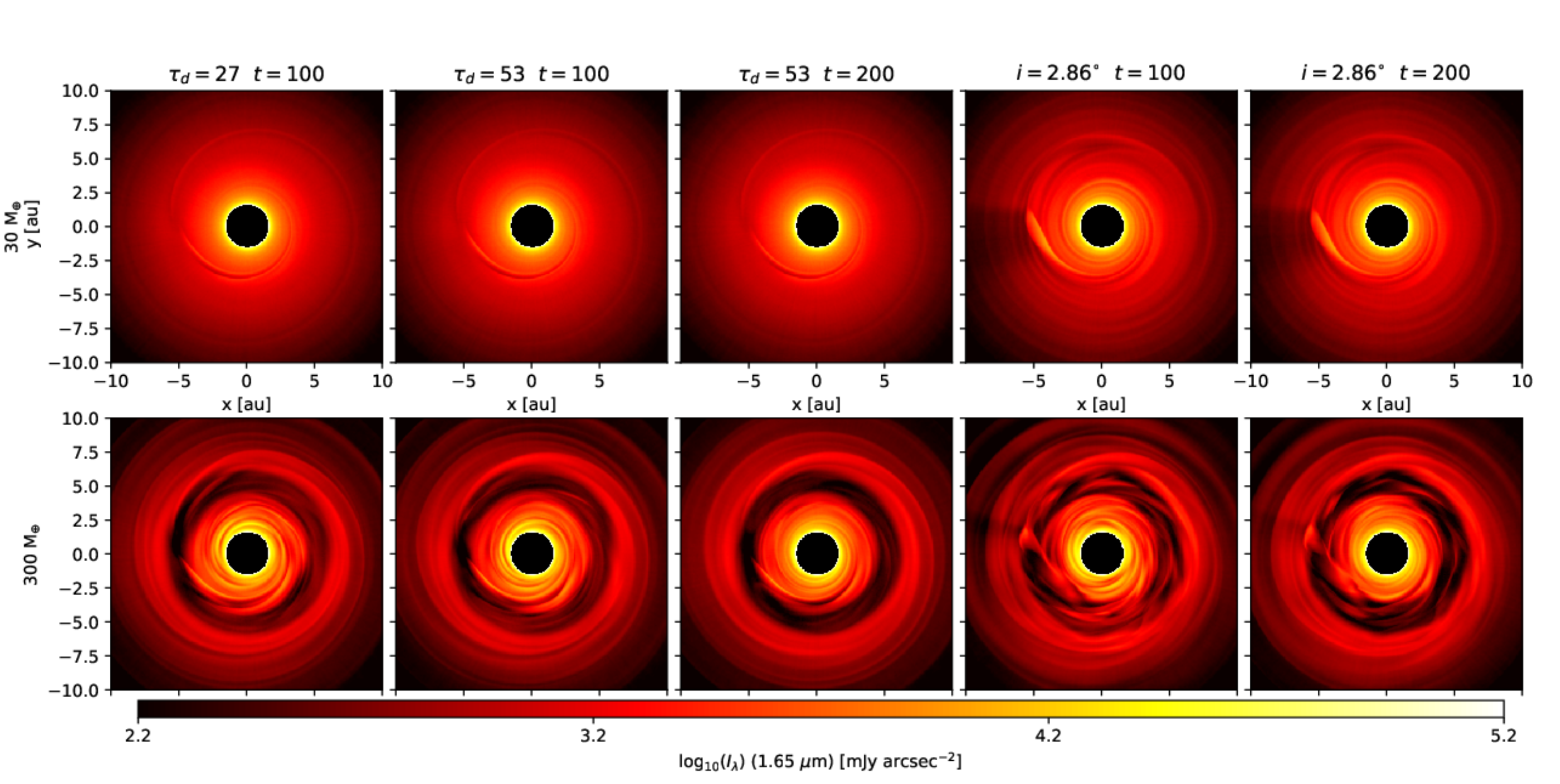}
\caption{Synthetic \emph{H-band} (1.65 $\mu$m) scattered light maps from 3D simulations 
involving 30 M$_{\earth}$ planets 
(Top) and 300 M$_{\earth}$ planets (Bottom) in which the planets' orbital inclination
angles have been damped compared to results from planets with a fixed orbit.  From
Left to Right: Damping time scale of $\tau_d = 27$ orbits at $t=100$ orbits
(planet inclination $i=0.1^{\circ}$), $\tau_d = 53$ at 
$t = 100$ orbits ($i=0.5^{\circ}$), $\tau_d = 53$ at $t = 200$ orbits ($i=0.08^{\circ}$), 
planet at a fixed orbit of $i=2.86^{\circ}$ at 
$t = 100$ orbits, and planet at a fixed orbit at $t = 200$ orbits. }\label{dampedI}
\end{figure*}

The synthetic scattered light results of these damped planetary orbits are presented
in Figure \ref{dampedI}, and compared to synthetic images of planets with 
sustained inclined orbits.  In the case of the 30 M$_{\earth}$ planet, the 
disc seems remarkably robust to the initial upper-atmospheric disruption 
due to the inclined planet.  There are little to no indications that these 
initial inclined orbits had any long-term effect on the structure of the disc 
for either damping time scales.  Once
the planet's orbit had become aligned with the disc midplane, the disc was able 
to fully recover from the initial non-antisymmetric disruptions, even though the 
sustained inclined planet caused massive disruptions after only 100 orbits.

This is also true even in the case of the more massive planet. Whereas the disc 
feature resembling a possible protoplanet reveals itself in the sustained 
incline simulations as early as 100 orbits, there is no hint of this disc 
feature in any of the damped inclined simulations.  In fact, the 
synthetic images of planets with damped orbits in Figure \ref{dampedI} 
much more closely resemble the $i=0^{\circ}$ 3D images in Figure \ref{scatMaps}
than the corresponding image of the sustained inclinations at 
$t = 100$ and $t=200$ orbits.  

It would seem that viscous protoplanetary discs are especially 
resilient to single-event or short term perturbations.
If interactions with the 
host disc cause the orbits of slightly inclined isolated planets 
to damp to the midplane, the disruptive effects appear to be 
negligible on disc time-scales.  If, however, a slightly inclined perturbation 
is sustained for any reason, it causes disruptions and features in near-IR observations 
which are not seen in surface density or midplane components of the disc.
There are various known mechanisms that are capable of causing orbital inclination, 
such as the Kozai mechanism \citep{1962AJ.....67..591K}, planet-planet scattering \citep[e.g.,][]{2008ApJ...686..580C}, or 
mis-alignment of the disc itself \citep{2018MNRAS.475.5618B}.  Even though simulations show that the orbital 
inclination of an isolated planet within a disc will damp towards the midplane, 
these simulations do not address what initially caused the inclination or whether 
that cause is capable of sustaining such an inclination. 

While small inclinations could be quickly damped in a disc there are several 
mechanisms that could keep the inclination to a certain level.
\cite{2017MNRAS.465.3175M} found that small planets may have their inclinations 
excited from disc-planet interaction when considering how the planet’s luminosity heats its surrounding. 
Also the damping time is connected to the disc mass and viscosity, which are quite uncertain. 
Our $\alpha$-viscosity is well within the wide range of observed 
values \citep{2017ApJ...837..163R}, and was set at $\alpha = 4\times 10^{-3}$ to help dampen out vertical shear 
instability (VSI).  This value was obtained primarily using simulations with
an aspect ratio of $H_0/R_0 = 0.05$ \citep{2013MNRAS.435.2610N}.  The time-scales
for damping orbital inclination were also determined using this ratio for a planet at 
5.2 au.  Extending the orbital radius out to 30 au significantly lowers the effective 
gas density, changes the aspect ratio to $H_0/R_0 \approx 0.07$.  Any of these changes could 
impeded the disc's ability to dampen the orbit of the planet.  

As it is still 
unknown whether the inclination of planetary orbits, such as those 
seen in our own solar system, must be produced after the disc has dissipated, 
planet-disc orbital misalignment should not be dismissed as a potential cause of observed 
features in protoplanetary discs, and the effects of sustained inclination 
on the dynamics and observable features of these discs should continue to be 
investigated.

\section{Summary \& Conclusions}\label{summary}

We ran two- and three-dimensional hydrodynamic simulations
of viscous, vertically-isothermal, protoplanetary discs with an 
embedded 30 M$_{\earth}$ planet and a 300 M$_{\earth}$ planet at 5.2 au.  
For all simulations we use a constant $\alpha$-viscosity parameter
of $\alpha = 4.0\times 10^{-3}$.
In the 3D cases we ran simulations in which the fixed orbit of the planets 
was aligned with the disc ($i=0^{\circ}$) and 
slightly misaligned 
with the disc ($i=2.86^{\circ}$), with and without damping the inclination over time.  
We post-processed each of these models using the Monte-Carlo radiation transfer code RADMC3D to created synthetic images which mimic how the features of these discs would be observed at effective 
\emph{H-band} wavelengths (i.e.,~1.65 \micron). \\*
\\*
Our results are summarized as follows:

\begin{enumerate}

\item In all models, the surface density evolution and the resultant features 
from the hydrodynamic simulations were effectively the same, regardless of whether the simulation was 2D or 3D,
and regardless of whether the orbit was slightly inclined. 
 
\item We found the $\tau_s = 2/3$ surface for an unperturbed disc to be a full 3.3 
thermal scale-heights above the midplane for our models, which means that observable features seen 
in \emph{H-band} scattered-light images primarily come from the gas/dust dynamics that occur in the 
upper-atmosphere of the disc, and not from surface density or midplane effects. 
This emphasizes the need for 3D models, especially for disc models used to explain scattered light observations.

\item  There are only slight differences in the features seen in scattered light images 
produced from the \verb+2D300+ and \verb+3D300A+ models, and their average intensity profiles from 
these images are quite similar. 
We conclude that 2D simulations are adequate to model observable features of discs under the limitation of an
embedded high-mass planet aligned with the disc midplane.

\item We find that inclined planets are able to ``churn up'' gas and dust material in the disc's upper-atmosphere.  
The results are strong perturbations in the scattered light intensity that even include shadow-causing features. 
Disc features caused by inclined planets can be 10-20 times as bright as their surrounding areas.
The model results also show dips in the average intensity profiles that could be falsely interpreted as the presence of a planet.

\item For a 300 M$_{\earth}$ planet which can clear out 
94\% of the material in its orbital path,
the vertical optical depth to the disc midplane for \emph{H-band} wavelengths 
is $\tau \approx 20$ at the orbital radius of the planet. 
We conclude that direct imaging of young planets embedded in the disc remains difficult to observe, even for massive planets in the gap.

\item A large-area shadowing of the disc (encompassing 30$^{\circ}$
of the disc in our case), can possibly be detected by current instruments, even it remains difficult to trace back to its physical origin.

\item Certain features from embedded planets with a sustained orbital incline produce 
distinct non-axisymetric disc features in scattered light when observed from above or below, which again shows the importance of modeling both hemispheres of the disc.   

\end{enumerate}

Finally, we emphasize that for a sustained inclination, or any effect 
which would cause a significant continuous disruption 
in the evolution of the disc's upper-atmosphere, 
fully 3D simulations are required to create 
scattered light models regardless of planet mass.

\appendix
\section{RADMC3D Setup}\label{RT_Setup}

To create synthetic scattered-light images from our hydrodynamic simulations 
as they would be observed in \emph{H-band} wavelengths ($1.65\micron$), we apply, in 
post-processing, the well-tested 
Monte Carlo radiative transfer code \verb+RADMC3D v0.41+ \citep{2012ascl.soft02015D} 
to our results.  We initially used the code to create a thermal structure of 
the dust embedded in the disc.  Preliminary experiments indicated a surface 
around 3--4 thermal scale-heights above the midplane.   
We can therefor reasonably ignore larger grains which settle
towards the midplane and assume a uniform dust-to-gas 
distribution of 0.01 for our purposes.

We model our radiative source as 
a star with a mass of $M_* = 1 M_{\sun}$, 
a radius of $R_* = 2.6 R_{\sun}$, an effective temperature 
of $T_* = 4280$ K, and assume isotropic scattering.  
As the code requires values in centimeter-gram-second (cgs) units,
we specified that the orbital radius of the planet was $R_0 = 5.2$ au,
and the surface density at that radius was $\Sigma_0 = 75$
g cm$^{-2}$.  This produces a 
disc with an initial mass of $M_d$ = 0.006 M$_{\sun}$. 
This corresponds to a 0.08 M$_{\sun}$ 
disc extended out to 100 au, consistent with observations 
\citep{2009ApJ...700.1502A} 
and other simulations of protoplanetary discs 
\citep{2012ApJ...761...95F, 2017ApJ...835...38D}.  We converted 
remaining values from the PLUTO simulations to cgs units using a 
unit length of $1 \ell_u = 5.2$ au, unit density $1 \rho_u = M_{\sun}/\ell_u^3$, 
and a unit velocity of $1 v_u = \sqrt{GM_{\sun}/\ell_u}/(2\pi)$, 
which corresponds to a unit time of one Keplerian orbit at 
$\mathbf{r} = (\ell_u, \pi/2, \phi)$.  

\begin{figure}
\includegraphics[width=\columnwidth]{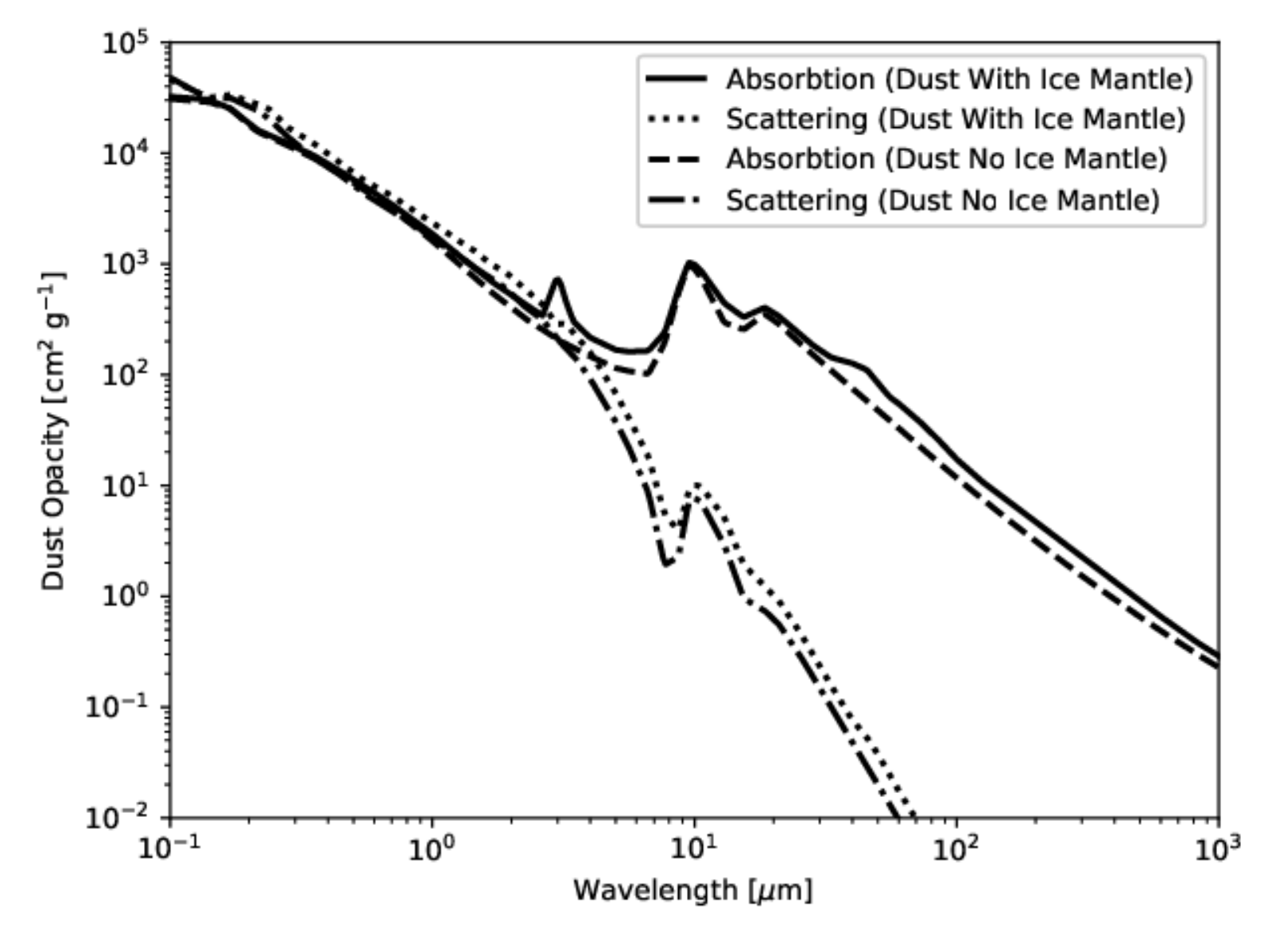}
\caption{$\log_{10}-\log_{10}$ plot of the wavelength 
dependent opacities 
from \citet{1993A&A...279..577P} used with RADMC3D
to create temperature 
profiles and scattered light images from our hydrodynamic simulations
of protoplanetary discs.}\label{opacities}
\end{figure}

For the dust opacity we used the wavelength-dependent table by 
\citet{1993A&A...279..577P}, 
which is based on a grain size distribution of 
amorphous carbon and silicate grains, and is especially suited
for the small grain sizes which are important for scattered-light
image observations.  For radiative transfer simulations 
on discs with an embedded planet at 5.2 au, we use opacity values 
for dust grains with no ice mantle, but for simulations scaled to 
approximate a simulation in which a planet is orbiting at 30 au, 
we use opacity values which include an ice mantle encasing
the dust grains to reflect the lower temperatures at this 
distance (see Figure \ref{opacities}).

Once the thermal structure of the disc has been created
we again use \verb+RADMC3D+ to create an image of our 
disc as it would be observed at 1.65 $\mu$m using second-order 
scattering, as well as calculating the height of the 
surface where the 
optical depth from an observer's point a view is $\tau_s = 2/3$.
 We refer to this as the 
\emph{observed surface} throughout this paper.

\section{Lower Dust-to-Gas Ratio}\label{lowDust}

We investigate the observable effect of diminished levels of small dust
grains in the disc's upper atmosphere by applying the \verb+RADMC3D+
code to the 3D simulations discussed in \S\ref{HD},
but using a dust-to-gas ratio of $10^{-4}$ 
(we omitted making new synthetic images 
from the 2D simulations).  The process 
of creating synthetic \emph{H-band} images is otherwise identical to
that described in Appendix \ref{RT_Setup}.     

\begin{figure}
\includegraphics[width=\columnwidth]{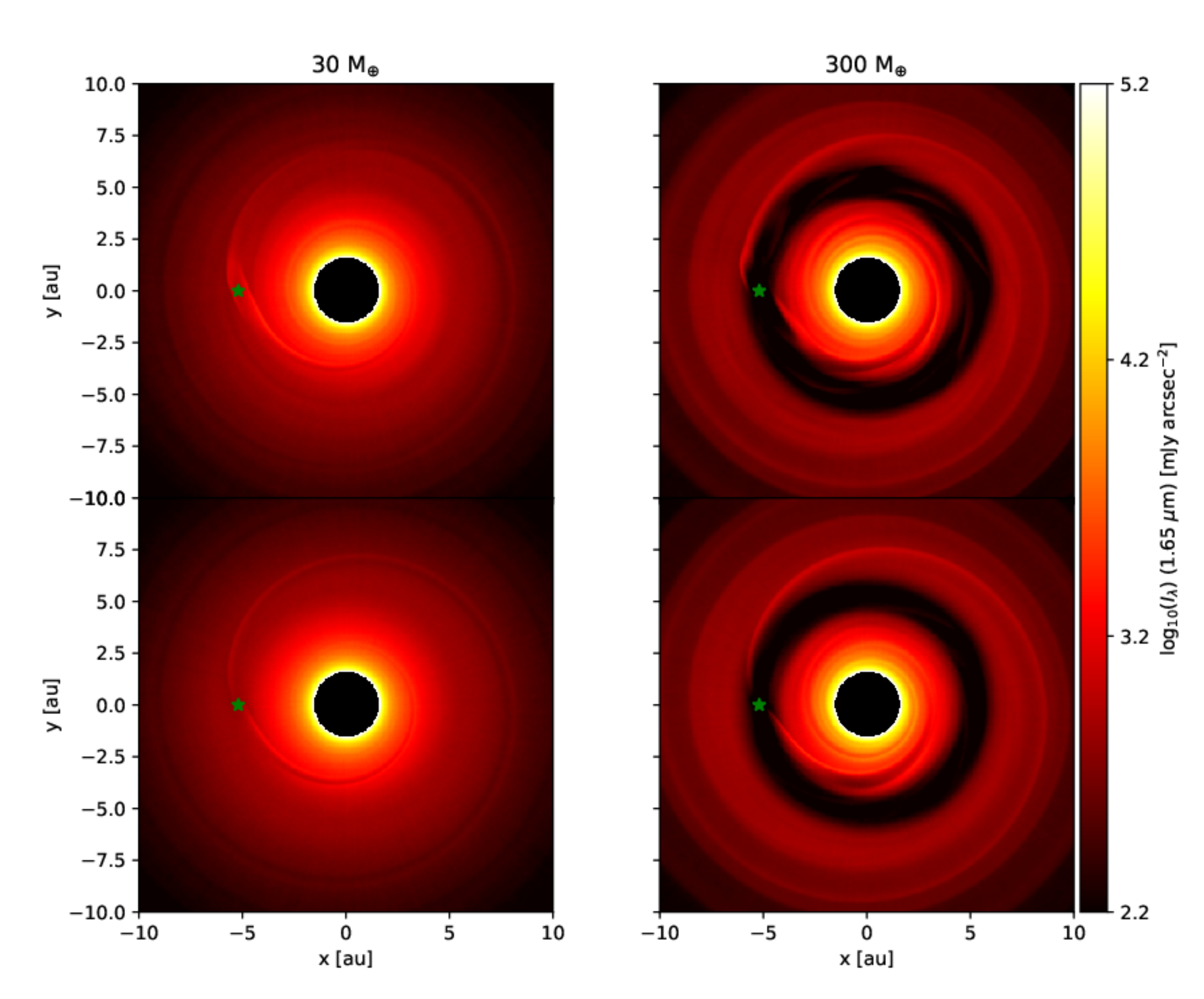}
\caption{Synthetic \emph{H-band} images from the 3D simulations involving 
30 M$_{\earth}$ planets (Left), 300 M$_{\earth}$ planets (Right)
with orbits with are aligned (Top) and mis-aligned (Bottom) with 
the disc, with the dust density reduced by a 
factor of 100 compared to those images seen in Figure \ref{scatMaps}.  
In this case the $\tau_s = 2/3$ surface of an unperturbed 
disc is $1.5H$ above the midplane at $R=R_0$.}\label{lowDenDust}
\end{figure}

When we applied this process using the new dust-to-gas 
ratio, we found that the $\tau_s = 2/3$ surface at $R=R_0$
for the unperturbed disc was only
$1.5H$ above the midplane.  Figure \ref{lowDenDust} shows 
the synthetic images produced from the 3D simulations 
involving 30 and 300 M$_{\earth}$ planets with inclined 
and aligned orbits.  These images reveal disc 
features which are much closer to the midplane, and the observable
disruptions from the inclined planets are not as apparent, 
and do not produce shadows (compare to Figure \ref{scatMaps}).   

\section*{acknowledgments}
We would like to thank Michael Brotherton at the University of Wyoming
for his feedback on the draft.

Mario Flock has received funding from the European Research Council 
(ERC) under the European Union's Horizon 2020 research and innovation
programme (grant agreement 757957).
  
All simulations were done using the Advanced Research Computing Center. 2012. 
Mount Moran: IBM System X cluster. 
Laramie, WY: University of Wyoming. http://n2t.net/ark:/85786/m4159c, 
and the Advanced Research Computing Center (2018) Teton Computing Environment, Intel x86\_64 cluster. 
University of Wyoming, Laramie, WY https://doi.org/10.15786/M2FY47

All plots were made using Matplotlib \citep{Hunter:2007}.
  
\bibliographystyle{mnras}
\bibliography{klosterV3Bib}

\end{document}